\documentclass[aip,preprint,cha,twocolumn,tightenlines,amsmath,amssymb,amsfonts,showpacs,10pt]{revtex4-1}

\usepackage[english]{babel}
\usepackage{times}
\usepackage{color}
\usepackage{oldgerm,amsmath,pifont,amsfonts,amssymb,latexsym}
\usepackage{float,xspace,calc,theorem,ifthen}
\usepackage{enumerate,psboxit,subfigure}
\usepackage[dvips]{epsfig,graphicx}
\usepackage{hyperref}
\hypersetup{
    colorlinks,%
    citecolor=blue,%
    filecolor=blue,%
    linkcolor=blue,%
    urlcolor=blue
}
\usepackage{changebar}
\usepackage{verbatim}

\newcommand{\rz}{{\mathbb R}}
\newcommand{\nz}{{\mathbb N}}
\newcommand{\zz}{{\mathbb Z}}

\begin{document}

\title{Dependence of chaotic diffusion on the size and position of holes}

\author{Georgie Knight}
\email{g.knight@qmul.ac.uk}
\affiliation{School of Mathematical Sciences, Queen Mary University of London, Mile End Road, London E1 4NS, UK}
\author{Orestis Georgiou}
\email{orestis@pks.mpg.de}
\affiliation{Max-Planck-Institut f\"{u}r Physik komplexer Systeme, N\"{o}thnitzer Stra{\ss}e 38, 01187 Dresden, Germany}
\author{Carl P. Dettmann}
\email{carl.dettmann@bristol.ac.uk}
\affiliation{School of Mathematics, University of Bristol, University Walk, Bristol BS8 1TW, UK}
\author{Rainer Klages}
\email{r.klages@qmul.ac.uk}
\affiliation{School of Mathematical Sciences, Queen Mary University of London, Mile End Road, London E1 4NS, UK}

\begin{abstract} A particle driven by deterministic chaos
and moving in a spatially extended environment can exhibit normal
diffusion, with its mean square displacement growing proportional to
the time.  Here we consider the dependence of the diffusion
coefficient on the size and the position of areas of phase space
linking spatial regions (`holes') in a class of simple
one-dimensional, periodically lifted maps. The parameter
dependent diffusion coefficient can be obtained analytically via a
Taylor-Green-Kubo formula in terms of a functional recursion
relation. We find that the diffusion coefficient varies
non-monotonically with the size of a hole and its position, which
implies that a diffusion coefficient can increase by making the hole
smaller. We derive analytic formulas for small holes in terms of
periodic orbits covered by the holes. The asymptotic regimes that we
observe show deviations from the standard stochastic random
walk approximation. The escape rate of the
corresponding open system is also calculated. The resulting parameter
dependencies are compared with the ones for the diffusion coefficient
and explained in terms of periodic orbits.  \end{abstract}


\maketitle

\begin{quotation} The diffusion coefficient of a physical system
involves a macroscopic measurement of the linear increase in the mean
square displacement of an ensemble of microscopic points. The source
of this increase is often modeled as the result of many random
collisions between the elements of a system, for example caused by
thermal motions. However, on a molecular level, the rules governing
the collisions are deterministic. In order to obtain a full
understanding of the process of diffusion, one must take these
deterministic rules into account. The discovery of chaotic dynamical
systems which exhibit diffusion has helped facilitate this, as one can
consider a relatively simple, low-dimensional setting in which
diffusion can be studied along with the deterministic rules at the
heart of diffusion. In this setting, one can consider how the
diffusion coefficient reacts to the variation of control parameters,
and hence gain understanding of more complicated diffusion
processes. Here we consider a simple, one-dimensional dynamical system
which consists of an interval map, periodically copied over the real
line with each interval connected to its neighbors via two regions
which we call holes.  We analytically derive the diffusion coefficient
for this system and study the effects of the size and position of the
holes.  We show that the position of a hole has a crucial impact on
the diffusion coefficient, sometimes dominating that of the size of a
hole. We further study the diffusion coefficient in the asymptotic
regime of small holes. We find analytic expressions for the diffusion
coefficient in terms of the limiting behavior of a hole, in particular
whether they contain a short periodic orbit. This allows construction
of a periodic orbit expansion for small but finite holes generalizing
simple random walk theory, which is based on an effectively
uncorrelated dynamics. Finally, we
compare our diffusion results with those obtained previously for the
escape rate and discuss the similarities and differences that arise.
\end{quotation}

\section{Introduction}
\label{sec:intro}

Recently there has been a surge of interest from mathematicians and
physicists on dynamical systems with holes, that is, subsets of phase
space that allow trajectories to leak out. Varying the
position of such holes in simple dynamical systems has exposed a
strong link between the average lifetime of chaotically transient
orbits and the location of short periodic orbits \cite{PaPa97}. Work on the escape properties of the doubling map and related
systems has revealed fundamental results for finite times, non-monotonic
dependence of escape rates on hole sizes, and a precise slowing of
escape for small holes containing a short periodic orbit \cite{Bun11}. The small hole effects have been
 generalized \cite{KeLi09,DeWr11,CPDett11}, and noise effects have also been considered \cite{Altmann2010,FaEc03}; for reviews see Refs.\onlinecite{AT,DY,DettCh}.
This work begs the
question: How do transport processes such as diffusion
depend upon the position and size of a hole? We answer this question
for the diffusion coefficient of an extended dynamical
system, interpreting the holes now as the links between
spatially separate regions of phase space. Diffusion is a
fundamental transport process of many-particle systems, the
study of which cross-links transport theory in statistical mechanics
with dynamical systems theory \cite{Do99,Gasp,Kla06,CAMTV01}.
The linear increase in time of the mean square displacement of an
ensemble of points in a chaotic, deterministic dynamical system is a
process known as `chaotic diffusion' or `deterministic diffusion'
\cite{SFK,GeNi82,GF2,Do99,RKdiss,Gasp,Kla06,CAMTV01}. Chaotic
diffusion provides a setting where the interaction between different
holes can be studied, an area that has yielded interesting results
regarding escape rates from circular \cite{BunDet05} and diamond
\cite{BunDet07} billiards with two holes, and very recently regarding
transmission and reflection rates in stadium billiards \cite{DeOr2011}
and the bouncer model \cite{LeDe11}. A general relation between escape
and diffusion has been established by the escape rate theory of
chaotic diffusion, which provides exact formulas expressing transport
coefficients in terms of escape rates in spatially extended systems
with absorbing boundary conditions \cite{GN,Gasp,RKdiss,Do99,Kla06}.

Much research has gone into studying the parameter dependence of the
diffusion coefficient in simple one-dimensional maps
\cite{SFK,GeNi82,GF2}. For low-dimensional, spatially periodic chaotic
dynamical systems the diffusion coefficient is often found to be a
fractal function of control parameters, exhibiting non-trivial fine
scale structure even in apparently simple examples
\cite{RKD,GaKl,Kla06}. The source of this fractality is typically
explained in terms of topological instability under parameter
variation of the underlying dynamics \cite{CAMTV01,Crist06}. However
there exist systems that display these hallmarks of fractality but
nevertheless have a linear diffusion coefficient
\cite{Kni11}. Therefore there is still work to be done explaining the
phenomenon of fractal diffusion coefficients in one dimension, let
alone attempting to answer questions about higher dimensional, more
physical systems like sawtooth maps \cite{DMP89}, standard maps
\cite{ReWi80} or particle billiards \cite{HaGa01,HaKlGa02} where
analytical results are lacking, as are answers to basic questions
about the structure of the diffusion coefficient. Previous work has
focused on deriving and understanding the diffusion coefficient under
smooth variation of control parameters of the dynamics
\cite{Kla06}. In this setting the reduced, modulo one dynamics of a
system will change with parameter variation \cite{CAMTV01}. Here we
switch focus and study a system where the reduced dynamics does not
change \cite{GaKl,Kni11,KnKl11b}. The main aim of this paper
is to see how the diffusion coefficient varies with the size and
position of a hole in a paradigmatic `toy model'.
In the process we find new relations between diffusion and escape rates,
between diffusion and periodic orbits, and we extend the small hole theory
of Refs.\onlinecite{KeLi09,Bun11} to diffusion.
What we learn here
can hopefully be transferred to more physically realistic dynamical
systems such as particle billiards \cite{Do99,Gasp,Kla06}.

In Sec.\ref{sec:sys} we define the dynamical system that will be the
main object of our study. It is a simple piecewise-linear chaotic map
of the real line which is a deterministic realization of a random
walk. It is constructed by copying and periodically lifting the
`Bernoulli shift' or `doubling map modulo one' over the whole real
line \cite{GaKl,Kni11,KnKl11b}. Modeling a coin tossing
process in terms of deterministic chaos, the doubling map modulo one
is one of the simplest dynamical systems displaying stochastic-like
properties making it an indispensable tool for studying statistical
mechanics in the setting of dynamical systems \cite{Do99,Ott}. In
addition, we choose the doubling map so that we can compare with the
results on escape rates from Refs.\onlinecite{KeLi09,Bun11} where the
doubling map modulo one was also focused upon. Furthermore,
the invariant measure of the doubling map modulo one is simply
Lebesgue, which helps make it amenable to analysis with the method we
will employ. The process of copying and periodically lifting a map is
the classical way to study chaotic diffusion in one dimension
\cite{SFK,GeNi82,GF2}. However we do not introduce diffusion into the
system through variation of a control parameter such as a shift or by
varying the slope. Rather we `dig holes' into the map that serve as
intervals where points can be iterated to a neighboring interval in
analogy with the work in Ref.\onlinecite{Bun11}.

We then analytically derive the diffusion coefficient as a
function of the size and position of a hole in this system via the
Taylor-Green-Kubo formula \cite{Do99,Kla06}, in terms of a functional
recursion relation. There are various methods for analytically
deriving diffusion coefficients \cite{CAMTV01,Crist06,Kla06} but the
method we use, developed in Refs.\onlinecite{RKdiss,GaKl,Kni11,KnKl11b}, is
the best suited to this setting. In Sec.\ref{sec:results} we look at
the analytical formulas derived in Sec.\ref{sec:sys} and find that the
diffusion coefficient varies as the position of a hole is varied, in
analogy with results on the escape rate. We also find that the
diffusion coefficient decreases non-monotonically as the size of a
hole decreases, a result that is different to the escape
rate. We explain this result in terms of the complicated
forward and back scattering rules associated with even simple
dynamical systems \cite{RKD,RKdiss,Kla06,Kni11}.

By using the same approach, similar results are obtained for
the diffusion coefficient of maps where the holes are not
placed symmetrically, and where the map generating the microscopic
dynamics is non-symmetric. Following this, we consider the case of
small hole size by deriving analytical expressions for the diffusion
coefficient which capture the asymptotic regime. We find that the
asymptotic regime is dependent upon the orbit structure of the
limiting point in an escape region, a result which goes beyond a
simple random walk approximation \cite{SFK,GF2,RKdiss,dcrc}. We
explain the results on position dependence, non-monotonicity and
asymptotic regimes by looking at the periodic orbit structure of the
map. Moreover, we build a periodic orbit expansion for small but
finite holes giving a more intuitive insight of the above. In
Sec.\ref{sec:escrate} we numerically calculate the escape rate for the
corresponding open system in order to compare with the
structure of the diffusion coefficient. We summarize our findings and
conclude our work in Sec.\ref{sec:conclusion}.

\section{Deriving the diffusion coefficient of a deterministic dynamical system.}
\label{sec:sys}

The dynamical system that we will study is based on the doubling map modulo one, whose phase space is simply the unit interval,
\begin{equation}
\tilde{M} (x)=
\left\{
\begin{array}{rl}
2x & 0\leq x <\frac{1}{2}\\
2x-1 & \frac{1}{2}\leq x < 1\end{array}\right. .
\label{Eq:Bern}
\end{equation}
The tilde in Eq.(\ref{Eq:Bern}) will be used throughout to signify a self-map. We turn Eq.(\ref{Eq:Bern}) into a dynamical system that exhibits diffusion in two steps. Firstly, we dig two symmetric holes into $\tilde{M}(x)$. Let $0\leq a_1 <a_2 \leq 1/2 \leq a_3 <a_4 \leq1$, with $a_4=1-a_1$ and $a_3=1-a_2$. For simplicity we let $h=a_2-a_1$ which is the size of a hole. We lift the map dynamics by $1$ for $x \in [a_1,a_2]$ and we lower the dynamics by $1$ for $x \in [a_3,a_4]$ to create a map $M(x):[0,1]\rightarrow [-1,2]$,
\begin{equation}
M (x)=
\left\{
\begin{array}{rl}
2x   & 0           \leq x <a_1\\
2x+1 & a_1         \leq x <a_2\\
2x   & a_2         \leq x <\frac{1}{2}\\
2x-1 & \frac{1}{2} \leq x <a_3\\
2x-2 & a_3         \leq x <a_4\\
2x-1 & a_4         \leq x \leq 1\end{array}\right. .
\label{Eq:Bern_lift}
\end{equation}
We label the intervals $I_L=[a_1,a_2]$ and $I_R=[a_3,a_4]$ for
convenience. We call $I_L$ and $I_R$ holes as they allow points
to escape from the unit interval to a neighboring interval.

Secondly, we periodically copy $M(x)$ over the entire real line with a lift of degree one such that,
\begin{equation}
M(x+b)=M(x)+b, \ \ b\in \zz,
\label{Eq:lift}
\end{equation}
so that $M(x):\rz\rightarrow \rz$. A uniform distribution of points on the unit interval will spread out when iterated under Eqs.(\ref{Eq:Bern_lift},\ref{Eq:lift}). The diffusion coefficient $D$, is defined as the linear increase in the mean square displacement of a distribution of points and is given by the Einstein formula in one dimension as
\begin{equation}
D= \lim_{n\rightarrow \infty} \frac{\left\langle \left( x_n-x_0\right)^2\right\rangle}{2n},
\label{Eq:ein_D}
\end{equation}
where $x_n$ is the position of a point $x_0$ at time $n$ which is given by $M^{n}(x_0)$ in the system we consider. The angular brackets represent an average over a distribution of points. In the setting we consider, this distribution is the invariant density of the system $\rho^*(x)=1$, and the average we interpret as an integral,
\begin{equation}
\left\langle ... \right\rangle = \int_0^1 ... \rho^*(x)dx.
\label{Eq:average}
\end{equation}
Eq.(\ref{Eq:ein_D}) can be rewritten in terms of the velocity autocorrelation function of the system as the Taylor-Green-Kubo formula \cite{Do99,Kla06},
\begin{equation}
D = \lim_{n\to\infty} \left(\sum_{k=0}^n \left\langle v_0(x) v_k(x) \right\rangle\right) -\frac{1}{2}\left\langle v_0(x)^2 \right\rangle\,,
\label{Eq:TGK}
\end{equation}
where $v_k(x)=\lfloor x_{k+1} \rfloor-\lfloor x_k \rfloor$ gives the integer value of the displacement of a point $x_0$ at time $k$. Considering Eq.(\ref{Eq:Bern_lift}) $v_k(x)$ takes the form,
\begin{equation}
v_k (x)=
\left\{
\begin{array}{rl}
0  & 0           \leq x_k <a_1\\
1  & a_1         \leq x_k <a_2\\
0  & a_2         \leq x_k <a_3\\
-1 & a_3         \leq x_k <a_4\\
0 &  a_4         \leq x_k \leq 1\end{array}\right. .
\label{Eq:vel}
\end{equation}
The leading order term of Eq.(\ref{Eq:TGK}), $D_{rw}$, is simply equal to
\begin{eqnarray}\nonumber
D_{rw}&=& \frac{1}{2}\int_0^1 v_0(x)^2 dx, \\
&=& \frac{(a_4-a_3)+(a_2-a_1)}{2} =h.
\label{Eq:Drw}
\end{eqnarray}
Eq.(\ref{Eq:Drw}) is the simple random walk result for diffusion that one obtains if higher order correlations are neglected \cite{SFK,GF2,RKdiss,dcrc}. In order to fully evaluate Eq.(\ref{Eq:TGK}) we define a recursive function $J^n(x):[0,1]\rightarrow \zz$ \cite{RKdiss,GaKl,Kni11,KnKl11b},
\begin{eqnarray}\nonumber
                J^n(x) &=& \sum_{k=0}^n  v_k(x)\\\nonumber
                       &=& v_0(x) +\sum_{k=0}^{n-1} v_k(\tilde{M}(x))\\
                       &=& v_0(x) +J^{n-1}(\tilde{M}(x)).
\label{Eq:jump}
\end{eqnarray}
We then define a cumulative function which integrates over Eq.(\ref{Eq:jump}) as in Eq.(\ref{Eq:TGK}),
\begin{equation}
                T(x)= \lim_{n\to\infty}T^n(x)=\int_0^xJ^n(y)dy.
\label{Eq:T}
\end{equation}
Due to the chaotic nature of the map $M(x)$, $J^n(x)$ will be a very complicated step function for high values of $n$, hence in the limit $n\rightarrow \infty$ $T(x)$ will be a fractal function exhibiting non-trivial fine scale structure \cite{RKdiss,Kla06,Kni11,Do99}. By combining Eq.(\ref{Eq:jump}) and Eq.(\ref{Eq:T}) we can solve $T(x)$ as a functional recursion relation. We use the conditions that $T(0)=T(1)=0$ and that the function $T(x)$ is continuous to obtain
\begin{equation}
T (x)=
\left\{
\begin{array}{lc}
\frac{1}{2}T (2x )                 & \ \ 0\leq x <a_1\\
\frac{1}{2}T (2x )   +x-a_1        & \ \ a_1\leq x <a_2\\
\frac{1}{2}T (2x )   +a_2-a_1      & \ \ a_2\leq x <\frac{1}{2}\\
\frac{1}{2}T (2x-1 ) +a_2-a_1      & \ \ \frac{1}{2}\leq x <a_3\\
\frac{1}{2}T (2x-1 ) +1-x-a_1       & \ \ a_3\leq x <a_4\\
\frac{1}{2}T (2x-1 )               & \ \ a_4\leq x \leq 1  \end{array}\right. \:.
\label{Eq:Tfull}
\end{equation}
Repeated application of the recurrence relation means we can solve Eq.(\ref{Eq:Tfull}) as an infinite sum,
\begin{equation}
                T(x)= \lim_{n\rightarrow \infty}\sum_{k=0}^n \frac{1}{2^k} t(\tilde{M}^k(x)),
\label{Eq:Tsum}
\end{equation}
where
\begin{equation}
t(x)=
\left\{
\begin{array}{lc}
0            & \ \ 0\leq x <a_1\\
x-a_1        & \ \ a_1\leq x <a_2\\
a_2-a_1      & \ \ a_2\leq x <a_3\\
1-x-a_1      & \ \ a_3\leq x <a_4\\
0            & \ \ a_4\leq x \leq 1  \end{array}\right. \:.
\label{Eq:tfull}
\end{equation}
Eq.(\ref{Eq:TGK}) can now be evaluated in terms of the functional recursion relation of Eq.(\ref{Eq:Tfull}) as
\begin{eqnarray}\nonumber
                D &=& \lim_{n\to\infty} \left(\int_0^1 v_0(x) \sum_{k=0}^n  v_k(x) dx\right) -\frac{1}{2}\int_0^1 v_0^2(x) dx \\\nonumber
                  &=& \lim_{n\to\infty}\left(\int_{a_1}^{a_2} J^n(x) dx -\int_{a_3}^{a_4} J^n(x) dx\right)-h\\
                  &=& T(a_2)-T(a_1)-T(a_4)+T(a_3)-h.
\label{Eq:D}
\end{eqnarray}
Finally, due to the condition that $I_L$ and $I_R$ are symmetrically positioned, $T(x)$ is a symmetric function. We can use this to simplify Eq.(\ref{Eq:D}) to
\begin{equation}
                D = 2T(a_2)-2T(a_1) - h.
\label{Eq:D2}
\end{equation}
Eqs.(\ref{Eq:Tsum},\ref{Eq:D2}) provide us with a very efficient way to evaluate the diffusion coefficient for any choice of position or size of $I_L$ and $I_R$. For a more detailed discussion of this method see Refs.\onlinecite{RKdiss,GaKl,Kla06,Kni11,KnKl11b}. We will evaluate Eq.(\ref{Eq:D2}) for a series of choices in the following section.

\section{Analyzing the diffusion coefficient}
\label{sec:results}

In this section we look at how the diffusion coefficient varies with
the position of the holes and the asymptotic behavior as the hole size
goes to zero.

\subsection{Position dependence}
\label{subsec:pos}

\begin{figure*}[htbp]
\begin{center}
\includegraphics[width=7.7cm]{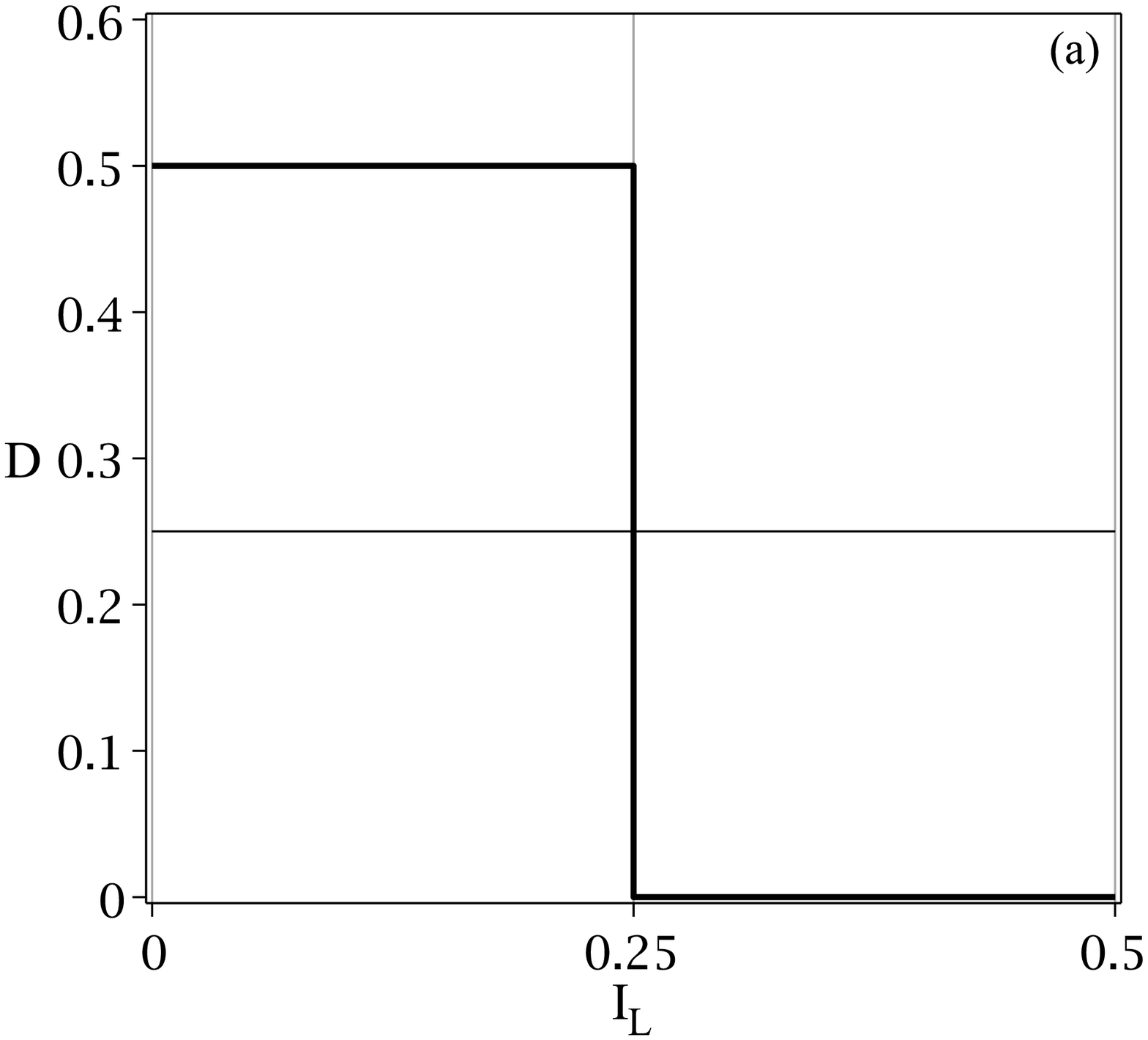} \includegraphics[width=7.7cm]{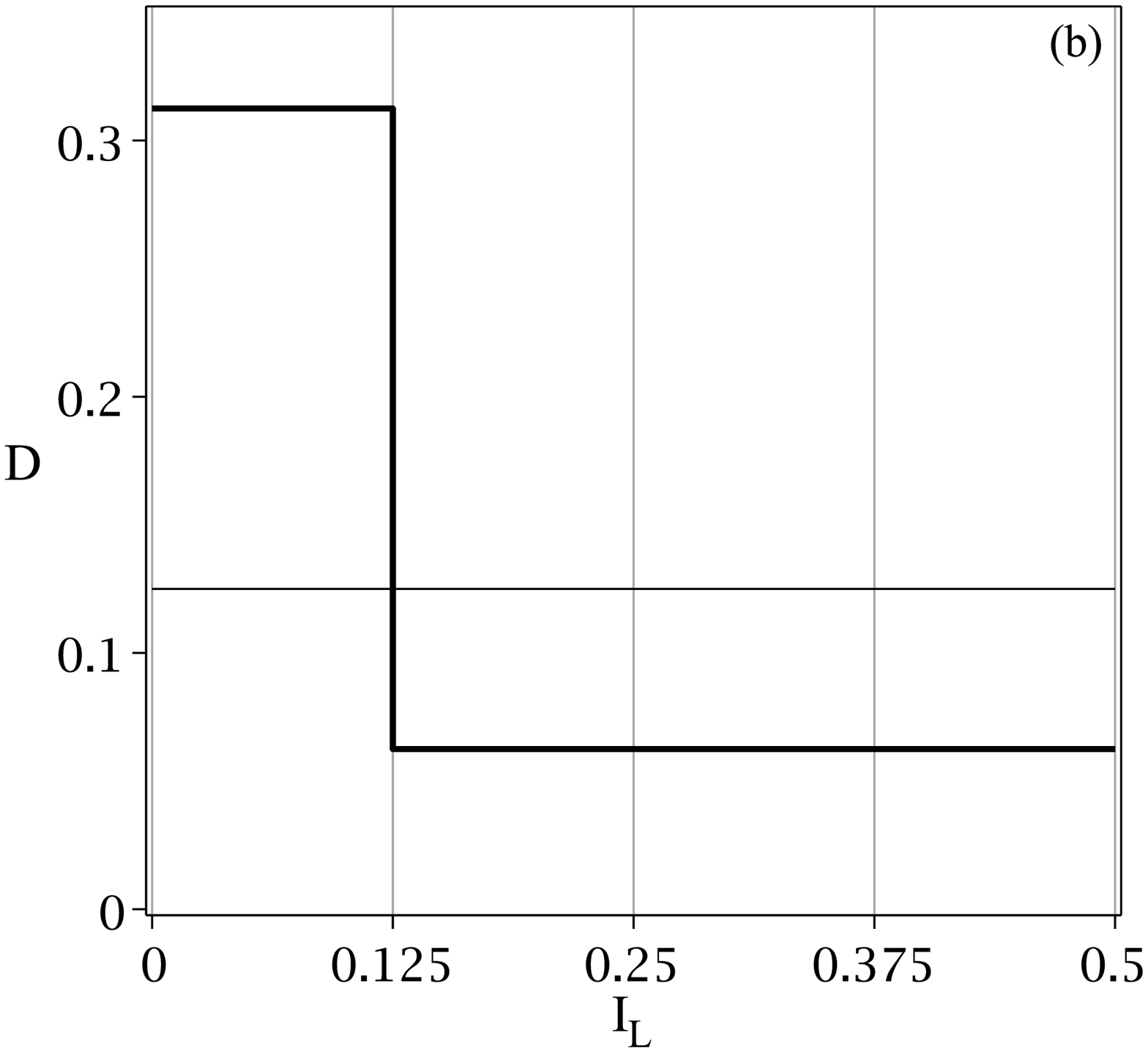}\\
\includegraphics[width=7.7cm]{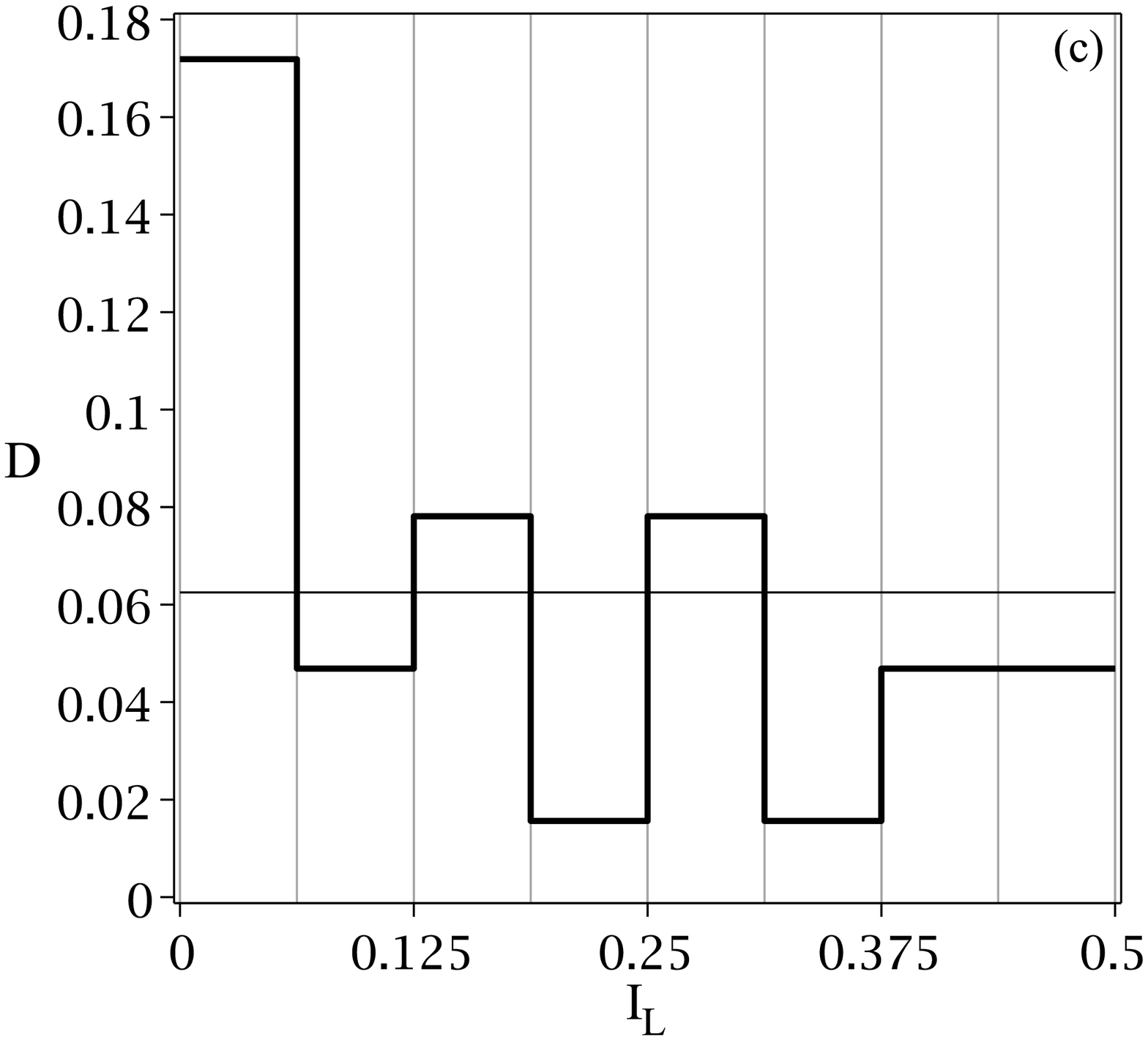} \includegraphics[width=7.7cm]{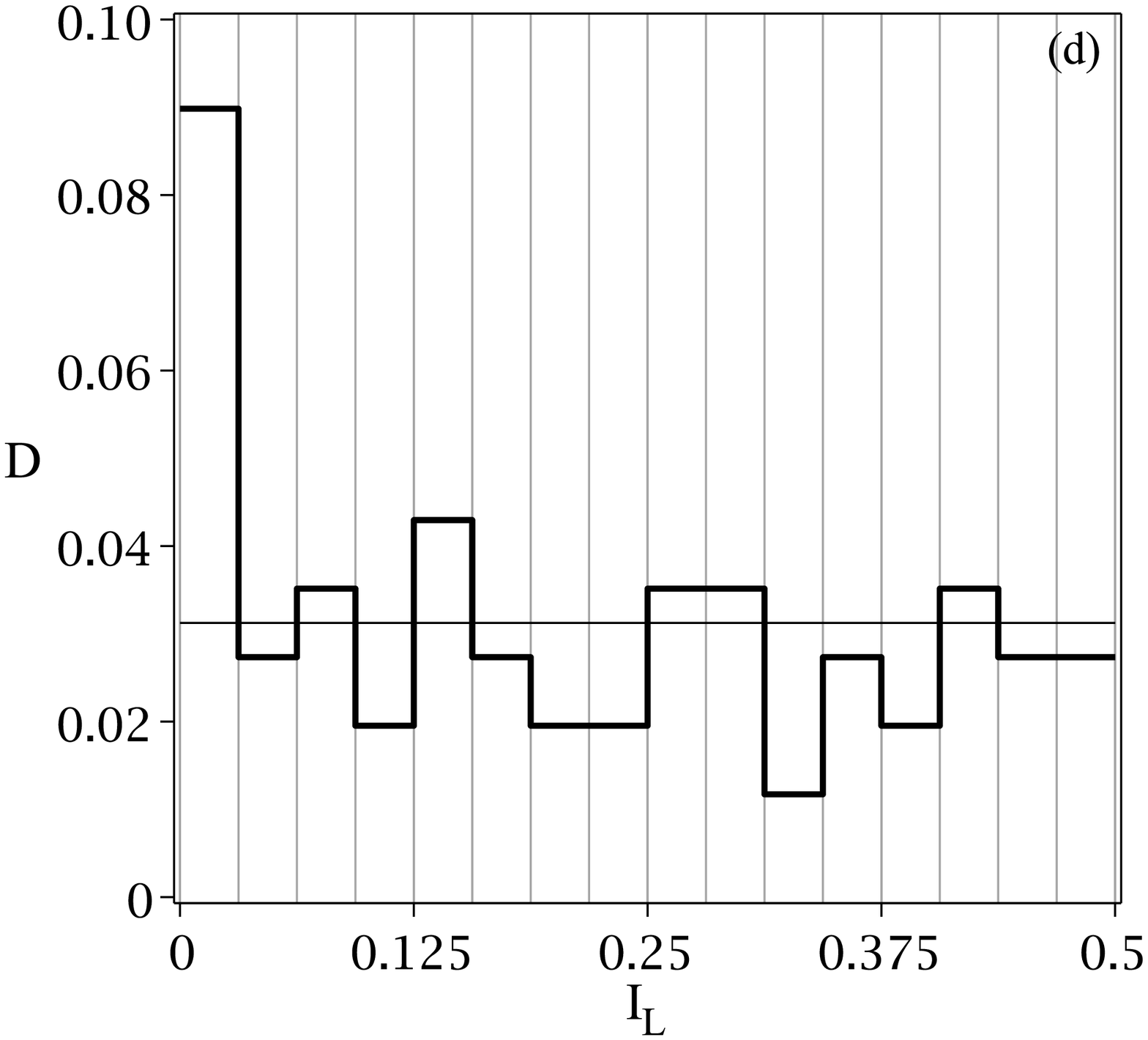}\\
\includegraphics[width=7.7cm]{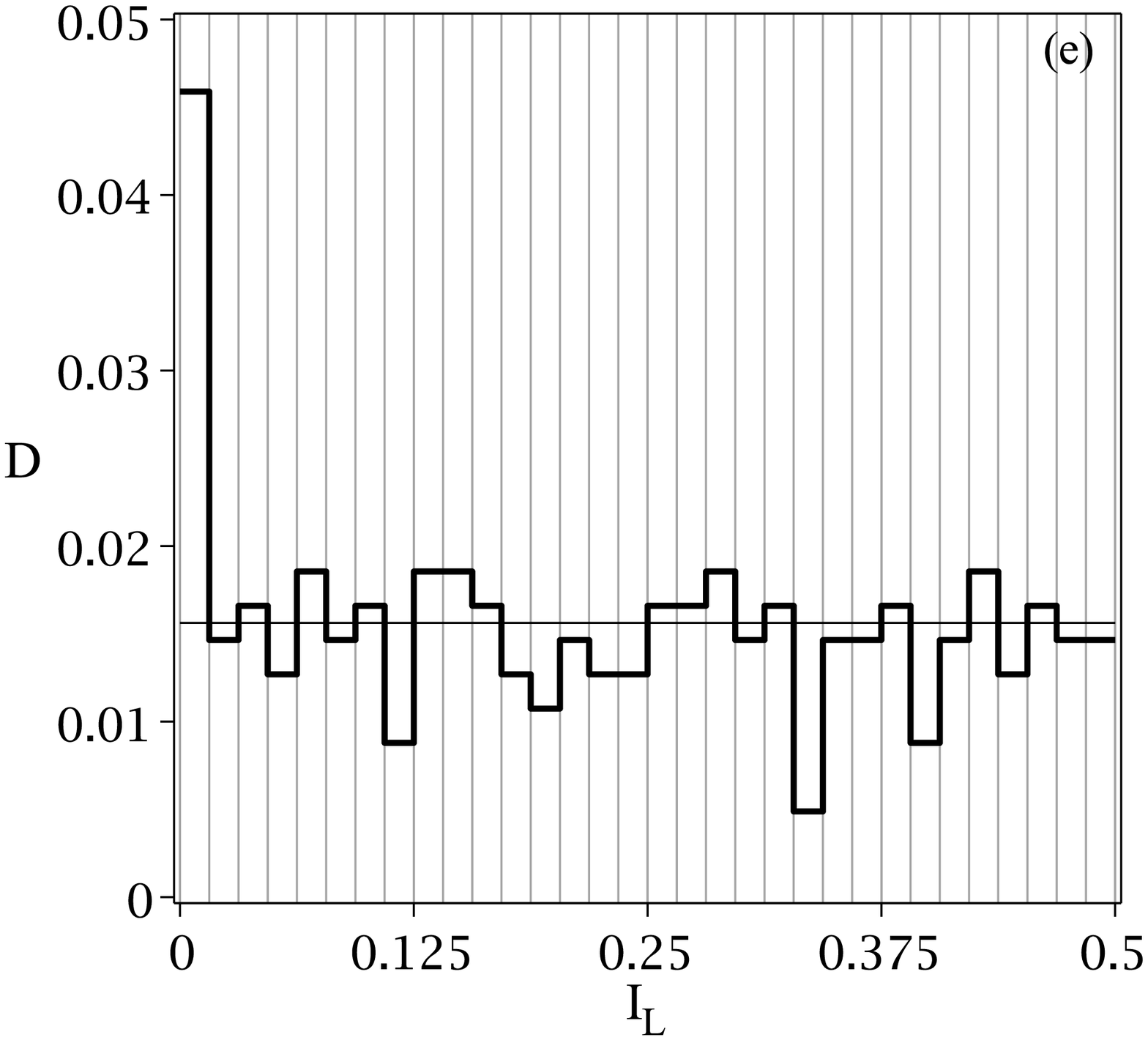} \includegraphics[width=7.7cm]{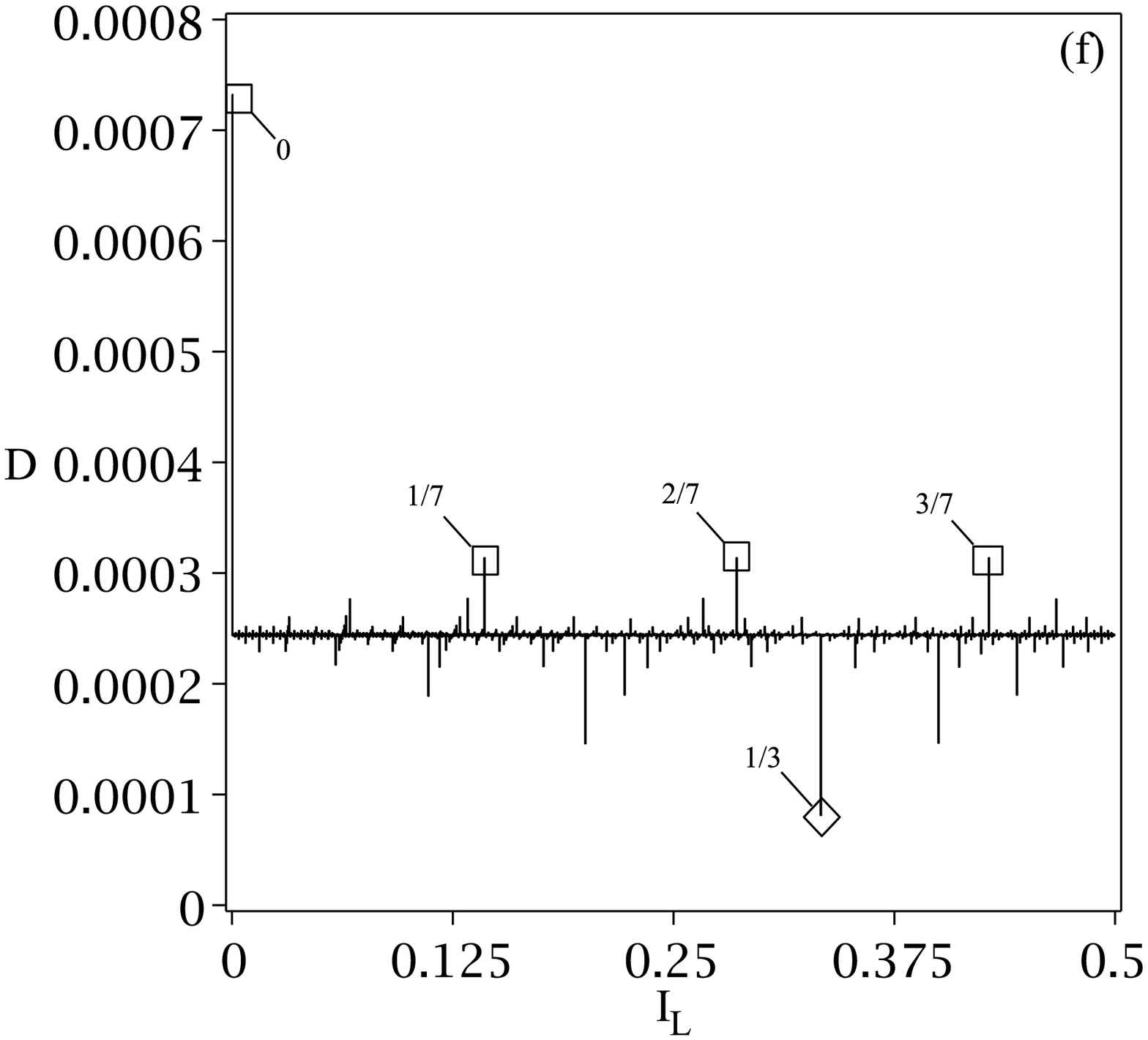}
\end{center}
\caption{The diffusion coefficients: In these figures
the diffusion coefficient $D$ is illustrated for the doubling map
$M(x)$ as a function of the position of the hole $I_L$ of size
$1/2^s$. In (a), (b), (c), (d), (e) and (f) it is $s=2,3,4,5,6$ and $12$
respectively. $D$ is given by the thick black
lines whilst the holes are highlighted by the thin vertical
lines. The thin horizontal lines are a guide to show the average value
$1/2^s$. The symbols in (f) refer to specific periodic orbits as discussed in the text.}
\label{fig:6D}
\end{figure*}

We evaluate Eq.(\ref{Eq:D2}) in a specialized setting where we restrict $I_L$ and $I_R$ to Markov intervals. That is, we choose the points $a_i, (i\in \{1,2,3,4\})$ to be dyadic rationals, i.e., for some fixed integer $s>0$ the points $a_i$ are of the form $r/2^s$ with $r \in \zz$ and $0\leq r \leq 2^s$. The points $a_i$ will then be pre-images of $1/2$ under the map $\tilde{M}(x)$. As $T(1/2)=a_2-a_1$, $T(a_i)$ can be evaluated with a finite sum rather than the infinite sum of Eq.(\ref{Eq:Tsum}). For each value of $s$ there are $2^{s-1}$ places to position an interval $I_L$ of size $2^{-s}$, with $I_R$ being determined by the symmetry condition. We can evaluate the diffusion coefficient at each of these choices via Eq.(\ref{Eq:D2}) and compare the results as the choices vary. For example, when $s=1$ there is only one choice for $I_L$, namely $a_1 =0, a_2=1/2$,
exactly corresponding to a simple random walk, with $1/2$ probability of moving left or
right at each step. The diffusion coefficient for this system is well known to be $1/2$, in
agreement with the more general expressions given here, Eq.(\ref{Eq:Drw}) for $h=1/2$ and
Eq.(\ref{Eq:D2}), as
\begin{eqnarray}\nonumber
               D &=& 2T(1/2)-2T(0)-(1/2)\\
                 &=& 1/2.
\label{Eq:D_ex}
\end{eqnarray}
For higher values of $s$ the diffusion coefficient varies with the position of the holes, see Fig.~\ref{fig:6D}. We see a step function that behaves increasingly erratically as the partition is refined and $s$ is increased.  We further note that the average of this step function can be calculated to be $\langle D_{s}\rangle=2^{-s}=h$ for a given $s$, which is the simple random walk solution of Eq.(\ref{Eq:Drw}).

The structure of the step functions in Fig.~\ref{fig:6D} can be explained in terms of the periodic orbits of the map $\tilde{M}(x)$ which correspond to   \textit{standing} or \textit{running} orbits of $M(x)$ \cite{KoKl03,Kla06,CAMTV01}. For example, if an image of $I_L$ overlaps $I_R$, one will find a lot of backscattering in the system, i.e., points that escape the unit interval via $I_L$ find themselves getting sent back via $I_R$ (and vice versa). This has the result of decreasing the diffusion coefficient relative to the random walk solution derived in Eq.(\ref{Eq:Drw}). In order to find intervals where this overlap occurs, we look for standing orbits by solving the simple equation $\tilde{M}^p(x)=x$ where $\tilde{M}^q(x)= 1-x$ for $q<p$. Due to the symmetry of the holes, the image of $I_L$ containing the solution of this equation will, after $q$ iterations, overlap with $I_R$ and backscattering will occur. The smaller values of $p$ will correspond to values of $x$ which give the most overlap and hence the most backscattering. For example,
\begin{equation}
               \tilde{M}(x)=1-x, \ \ x\in[0,1/2],\ \ \Rightarrow x =\frac{1}{3}.
\label{Eq:Dmin}
\end{equation}
Therefore if one places $I_L$ so that $x=1/3$ is in its interior, one will find the system has a relatively small diffusion coefficient due to the backscattering. This phenomenon caused by \textit{standing} orbits \cite{KoKl03,Kla06} is highlighted in Fig.~\ref{fig:6D}.

Alternatively, if the image of $I_L$ overlaps with itself consistently then one will find a higher diffusion coefficient. This is due to the presence of \textit{running} orbits or accelerator modes \cite{KoKl03,Kla06,CAMTV01} in such a system. In order to find such orbits, we solve the simple equation $\tilde{M}^p(x)= x$ where $\tilde{M}^q(x)\neq 1-x$ for $q<p$. For example $p=1$ gives
\begin{equation}
               \tilde{M}(x)=x, \ \ x\in[0,1/2],\ \ \Rightarrow x=0,
\label{Eq:Dmax1}
\end{equation}
and we can see in Fig.~\ref{fig:6D} that when the hole contains the point $x=0$ one has a high diffusion coefficient relative to the simple random walk result. When $p=2$,
\begin{equation}
               \tilde{M}^2(x)=x, \ \ \Rightarrow x=0,\frac{1}{3}.
\label{Eq:Dmax2}
\end{equation}
we can immediately throw the solution $x=1/3$ away as this result corresponds to $\tilde{M}(x)= 1-x$. However, for $p=3$
\begin{equation}
               \tilde{M}^3(x)=x, \ \ \Rightarrow x=0,\frac{1}{7}, \frac{2}{7},\frac{3}{7},
\label{Eq:Dmax3}
\end{equation}
and again we see in Fig.~\ref{fig:6D} that these values correspond to relatively high diffusion coefficients when they are in the interior of $I_L$. This process of pinpointing standing and running periodic orbits can be continued for higher iterations with relative ease as we are dealing with a full shift map and there is no need to prune any solutions. This technique helps explain the increasingly complicated step function that one obtains as $s$ is increased.

At first sight, the step functions illustrated in Fig.~\ref{fig:6D} do not appear to contain much interesting structure. However, upon closer inspection we notice that every `parent' hole of size $2^{-s}$ and associated diffusion coefficient $D_{s}$ splits into two `child' holes of size $2^{-(s+1)}$ and associated diffusion coefficients $D_{s+1}^{0}$ and $D_{s+1}^{1}$ respectively, such that
\begin{eqnarray}
                D_{s} &=& 2 D_{s+1}^{0} + 2 D_{s+1}^{1} - 2^{-s}.
\label{Eq:parent}
\end{eqnarray}
where superscripts $0,1$ correspond to left and right child hole respectively. To see this, one first needs to define cumulative functions $T^{0}(x)$ and $T^{1}(x)$ for the respective left and right child holes. Now since the cumulative functions are additive with respect to the holes we have that $T(a_{i})=T^{0}(a_{i})+T^{1}(a_{i})$ for $i=1\ldots4$. Moreover, since the
iterate of the parent hole endpoint $a_{i}$ always avoids both parent and child holes then $T^{0}(a_{i})=T^{1}(a_{i})$. Finally, considering the midpoint $a_{m}=(a_{2}+a_{1})/2$ of the parent hole which is also the right and left endpoint of the left and right child holes respectively, it follows from Eq.(\ref{Eq:Tfull}) that $T^{0}(a_m)-T^{1}(a_m)= (a_{2}-a_{1})/2$. Eq.(\ref{Eq:parent}) follows after expanding in terms of $T^{0}$ and $T^{1}$ and substituting the above relations.
Notice that recursive iteration of Eq.(\ref{Eq:parent}) $n$ times gives an expression for $D_{s}$ in terms of the $2^n$ child diffusion coefficients
\begin{equation}
D_{s}= (1-2^{n})2^{-s}+ 2^{n} \sum_{j\in\{0,1\}^n} D_{s+n}^{j},
\label{Eq:parent2}
\end{equation}
where the sum runs over all $2^n$ binary permutations of length $n$.  Rearranging this we find
\begin{equation}
\frac{D_s-2^{-s}}{2^{-s}}=\sum_{j\in\{0,1\}^n}\frac{D_{s+n}^j-2^{-s-n}}{2^{-s-n}}
\end{equation}
that is, the relative deviation of each diffusion coefficient from its mean is exactly additive.

\begin{figure*}[htb]
\begin{center}
\includegraphics[width=8.5cm]{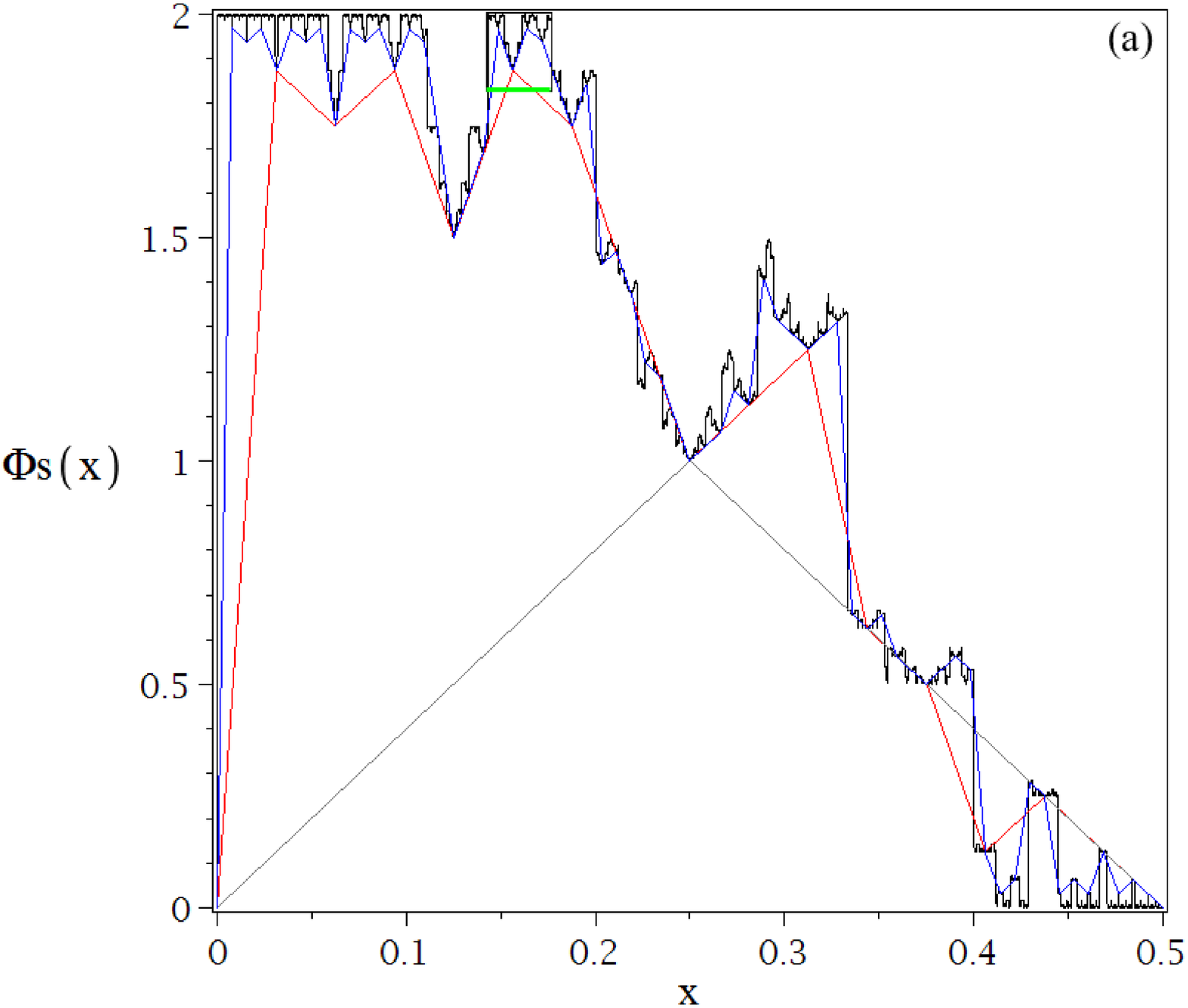} \includegraphics[width=8.5cm]{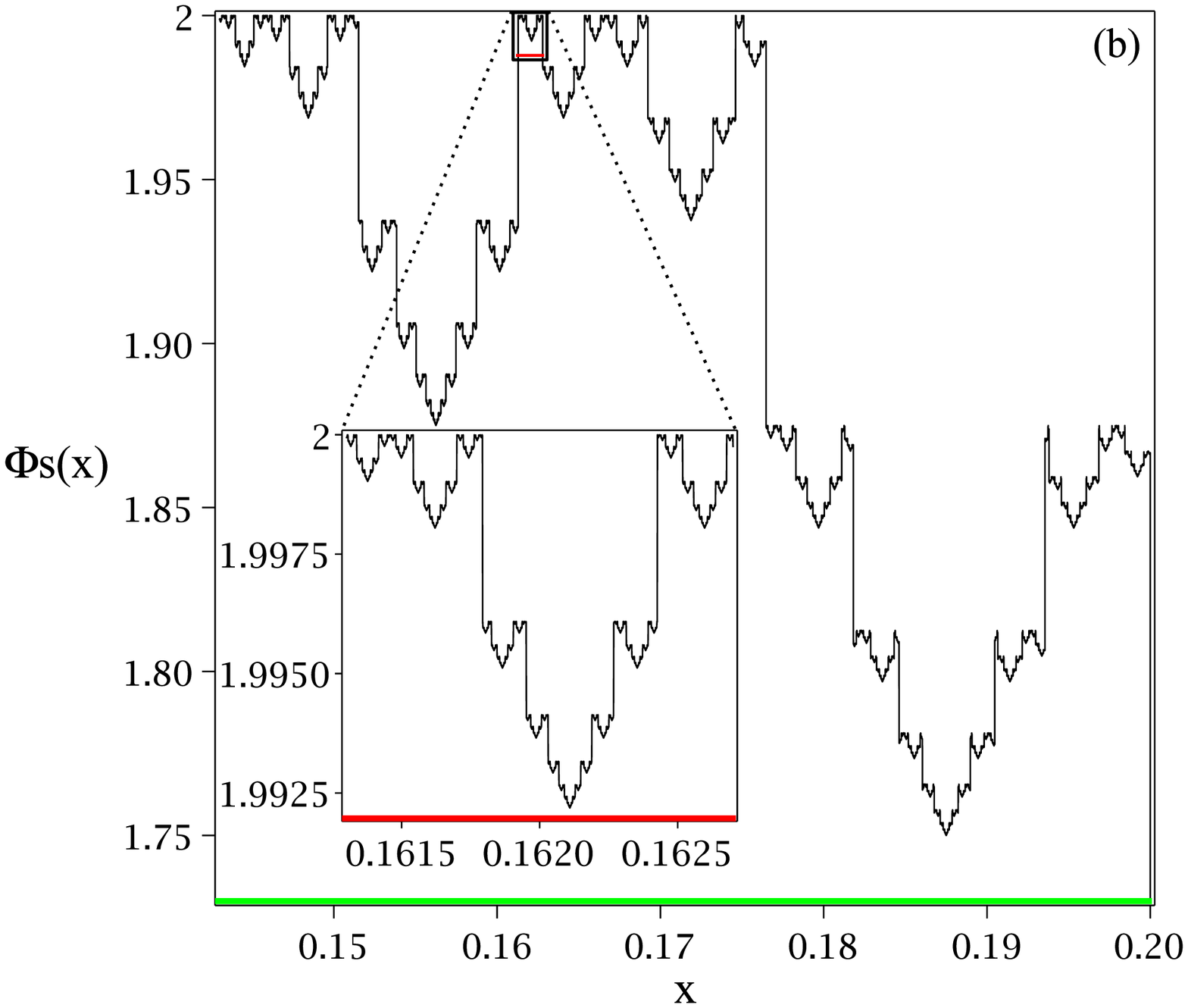}
\end{center}
\caption{(Color online) Cumulative integral function $\Phi_s(x)$: In this figure the self similarity and scaling that one sees by integrating over the position dependent diffusion coefficient for the doubling map is illustrated. In (a) the structure is seen emerging as the hole size $1/2^s$ is decreased. From $(0.1,0)$ upwards, $s=2$ (grey), $s=5$ (red), $s=8$ (blue) and $s=20$ (black). In (b) the region highlighted in (a) is blown up whilst the inset shows the highlighted region in (b) blown up in order to illustrate the self similarity and non-trivial fine-scale structure of the diffusion coefficient.}
\label{fig:cumul}
\end{figure*}

The above scaling and self-similarity structure (often considered
to be properties of fractal structures) of the step functions
illustrated in Fig.~\ref{fig:6D} can be further investigated by
defining a set of continuous, cumulative functions which integrate
over the step function, in the same way that $T(x)$ integrates over
the step function $J^n(x)$. In order to define such a function,
$\Phi_s(x)$, for a given $s$, we firstly subtract the average diffusion
coefficient $\left\langle  D_s \right\rangle=1/2^s$, and integrate over the resulting
step function. We then normalize this integral by multiplying it with
$2^{s+1}$ so that it can be easily compared with other values of $s$. Let
\begin{equation}
                 \Phi_s(x)  = 2^{s+1}\int_0^x \left(D(y)-2^{-s}\right) \ \ dy,
\label{Eq:cumul}
\end{equation}
where $D(y)$ refers to the diffusion coefficient of the dyadic interval
$I_L$ containing $y$. The solution to Eq.(\ref{Eq:cumul}) is illustrated
for several examples of $s$ in Fig.~\ref{fig:cumul}. We see that as $s$
increases, Eq.(\ref{Eq:cumul}) becomes a fractal function exhibiting
non-trivial fine scale structure and regions of scaling and self-similarity.
This structure is symptomatic of the dense set of periodic orbits which
exists in $\tilde{M}(x)$. In the limit of $s$ going to infinity, each periodic
orbit makes the diffusion coefficient deviate from the average hence one
obtains a dense step function. When this function is integrated over one sees
a function that contains a dense set of maxima and minima, hence a fractal.

Another interesting feature that we find in this system is that
reducing the size of the holes can sometimes have no effect on the
diffusion coefficient. As we calculated in Eq.(\ref{Eq:D_ex}), if the
holes are $I_L=[0,0.5]$ and $I_R=[0.5,1]$, the diffusion coefficient
is equal to $0.5$. However, we can reduce the hole so that
$I_L=[0,0.25]$ and $I_R=[0.75,1]$ and the diffusion coefficient
remains equal to $0.5$ as illustrated in Fig.~\ref{fig:6D}.(a). We
also see that if the holes are $I_L=[0.25,0.5]$ and $I_R=[0.5,0.75]$,
the diffusion coefficient is equal to $0$. This is due to a simple
trapping mechanism in which no diffusion occurs. In addition, reducing
the size of the holes can result in an increase of the diffusion
coefficient, i.e., the diffusion coefficient decreases
non-monotonically in some regions as the size of the holes is
decreased. One can check this by comparing the figures in
Fig.~\ref{fig:6D}. This feature is due to the fact that
increasing the size of the hole can in some cases introduce more
backscattering into the system thereby reducing the diffusion
coefficient: Consider the case where $I_L=[0.125,0.25]$ with diffusion
coefficient equal to $1/16$, Fig.~\ref{fig:6D}.(b), and compare with
the case with smaller holes, $I_L=[0.125,0.1875]$ but larger diffusion
coefficient $5/64$, Fig.~\ref{fig:6D}.(c). The dominant periodic
orbit (the one with the lowest period) in the interval
$[0.125,0.1875]$ is the orbit of the point $1/7$ which corresponds to
a running orbit as can be seen in Fig.~\ref{fig:6D}.(f).  The dominant
periodic orbit in $[0.1875,0.25]$ meanwhile is the orbit of the point
$1/5$ which corresponds to a standing orbit as can be seen in
Fig.~\ref{fig:6D}.(f).  The effect of decreasing the hole from
$[0.125,0.25]$ to $[0.125, 0.1875]$ is to remove the standing orbit of
$1/5$ and the backscattering associated with it.

One can also observe this phenomenon by looking at the fractal
structure illustrated in Fig.~\ref{fig:asymp}.(b). While $a_1=1/3$ is
fixed, the various maxima and minima that we see can be explained by
looking at the orbit of the point $a_2$. We see that when $a_2=5/12$
(corresponding to $h=1/12$), the orbit of $a_2$ is a standing orbit
and hence we see a striking minimum in the diffusion coefficient. If
we reduce $h$ so that $h=1/15$ with $a_2=2/5$, the orbit of $a_2$ is
now a running orbit and we observe a maximum in the diffusion
coefficient. These points are highlighted in
Fig.~\ref{fig:asymp}.(b). This explanation in terms of topological
instability under parameter variation is discussed further in
Refs.\onlinecite{RKD,RKdiss,Kla06,Kni11}.

\subsection{Non-symmetric models}
\label{Subsec:other}


\begin{figure*}[htbp]
\begin{center}
\includegraphics[width=5.9cm]{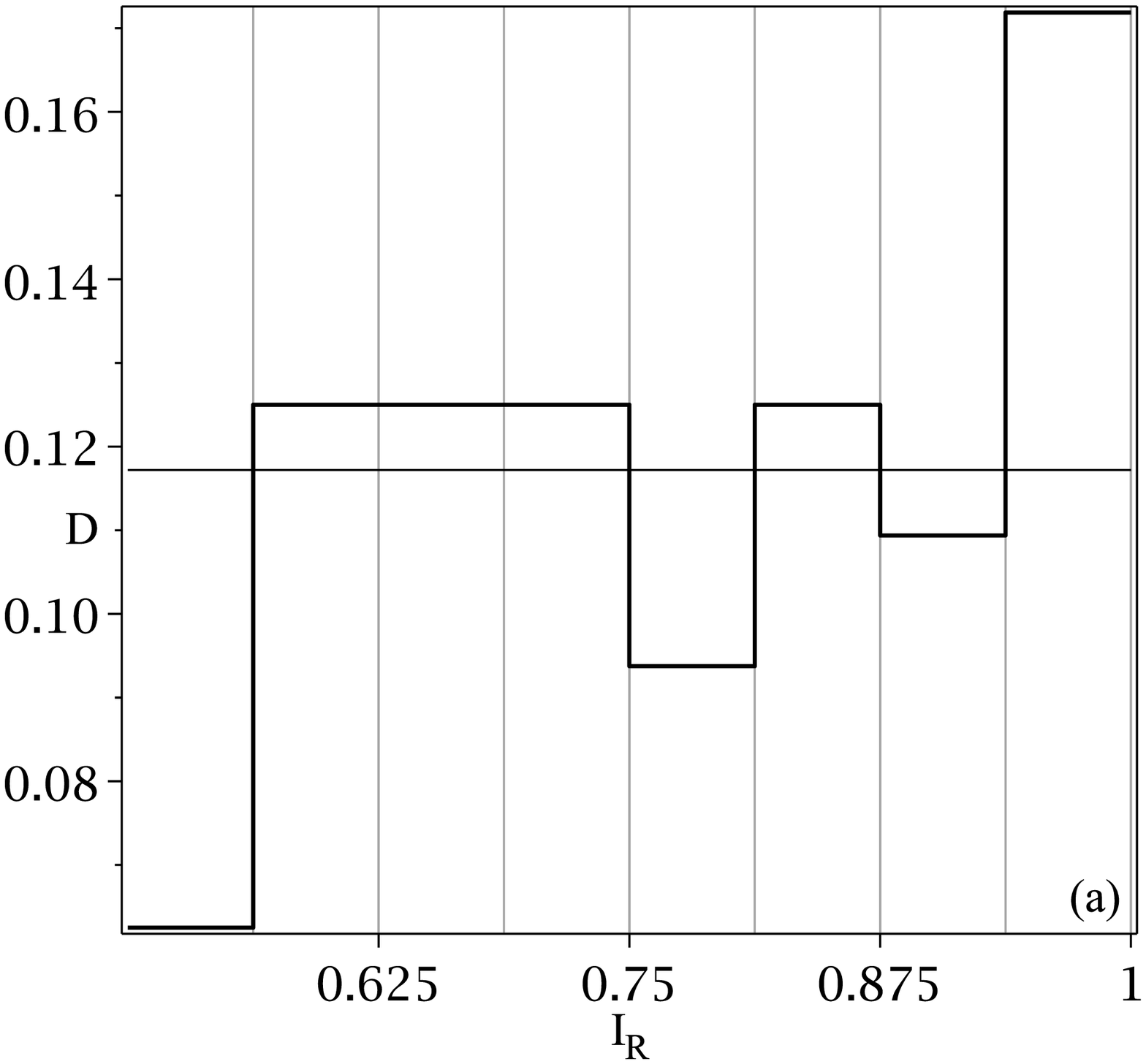} \includegraphics[width=5.9cm]{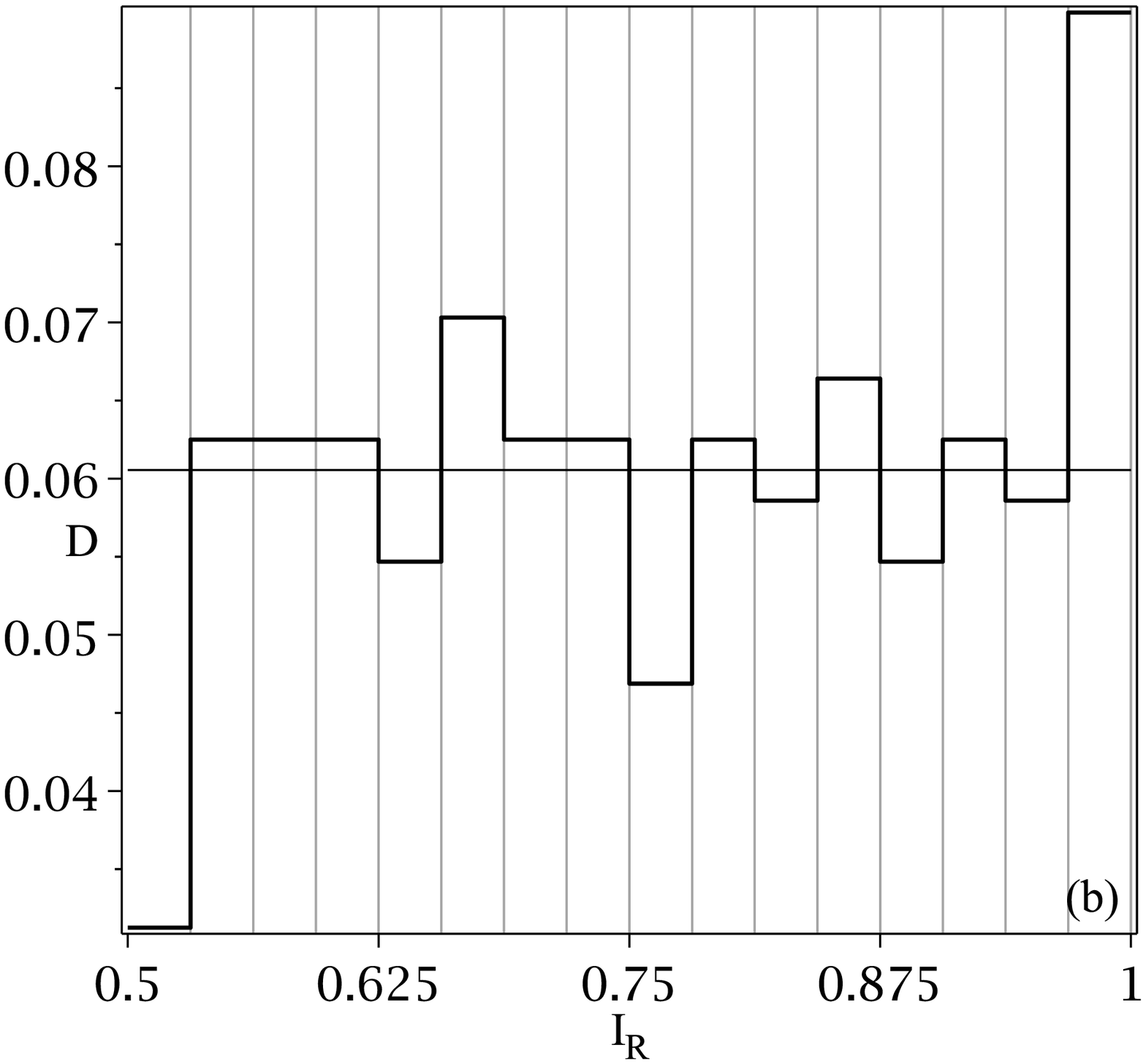} \includegraphics[width=5.9cm]{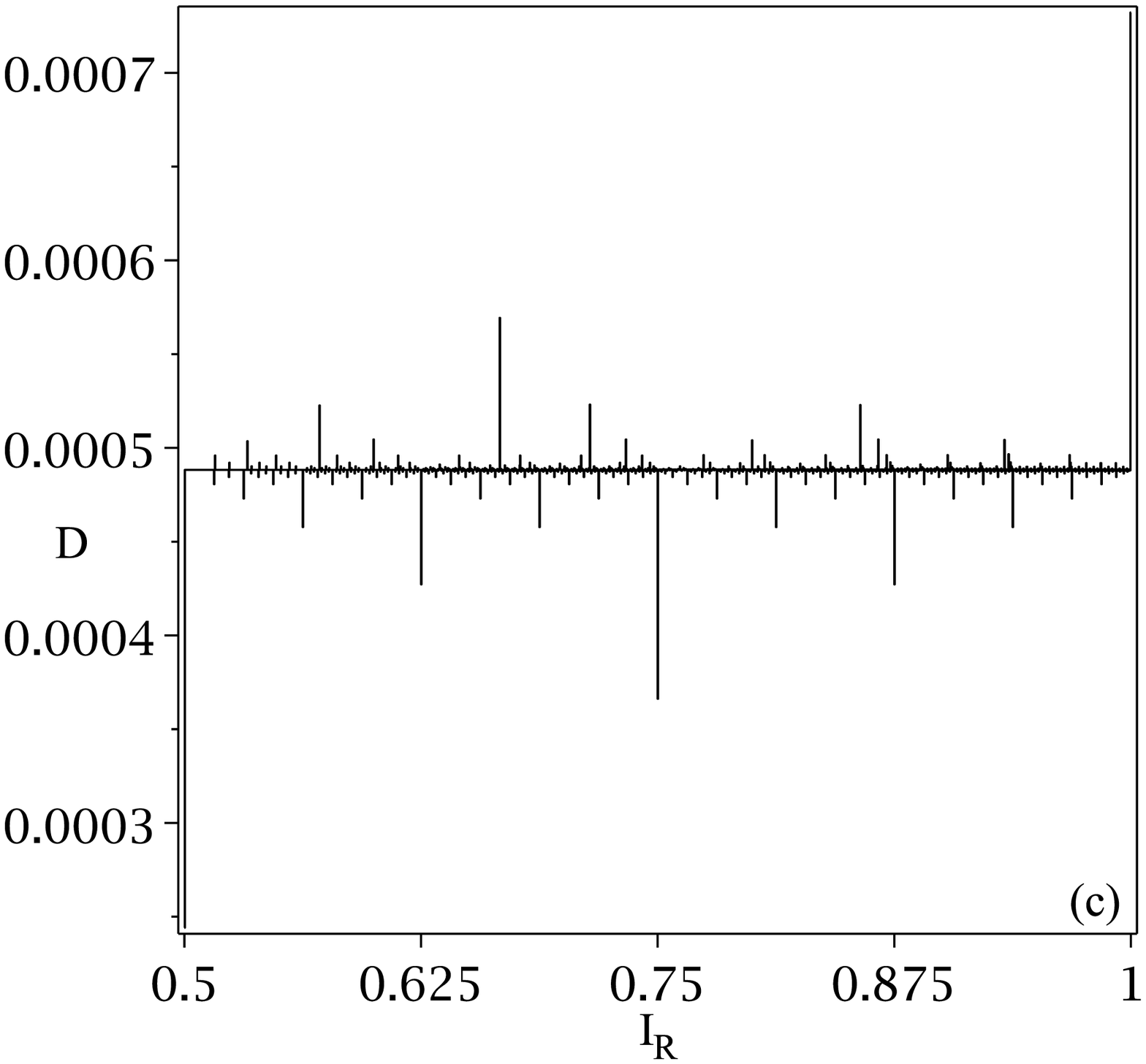}\\
\includegraphics[width=5.9cm]{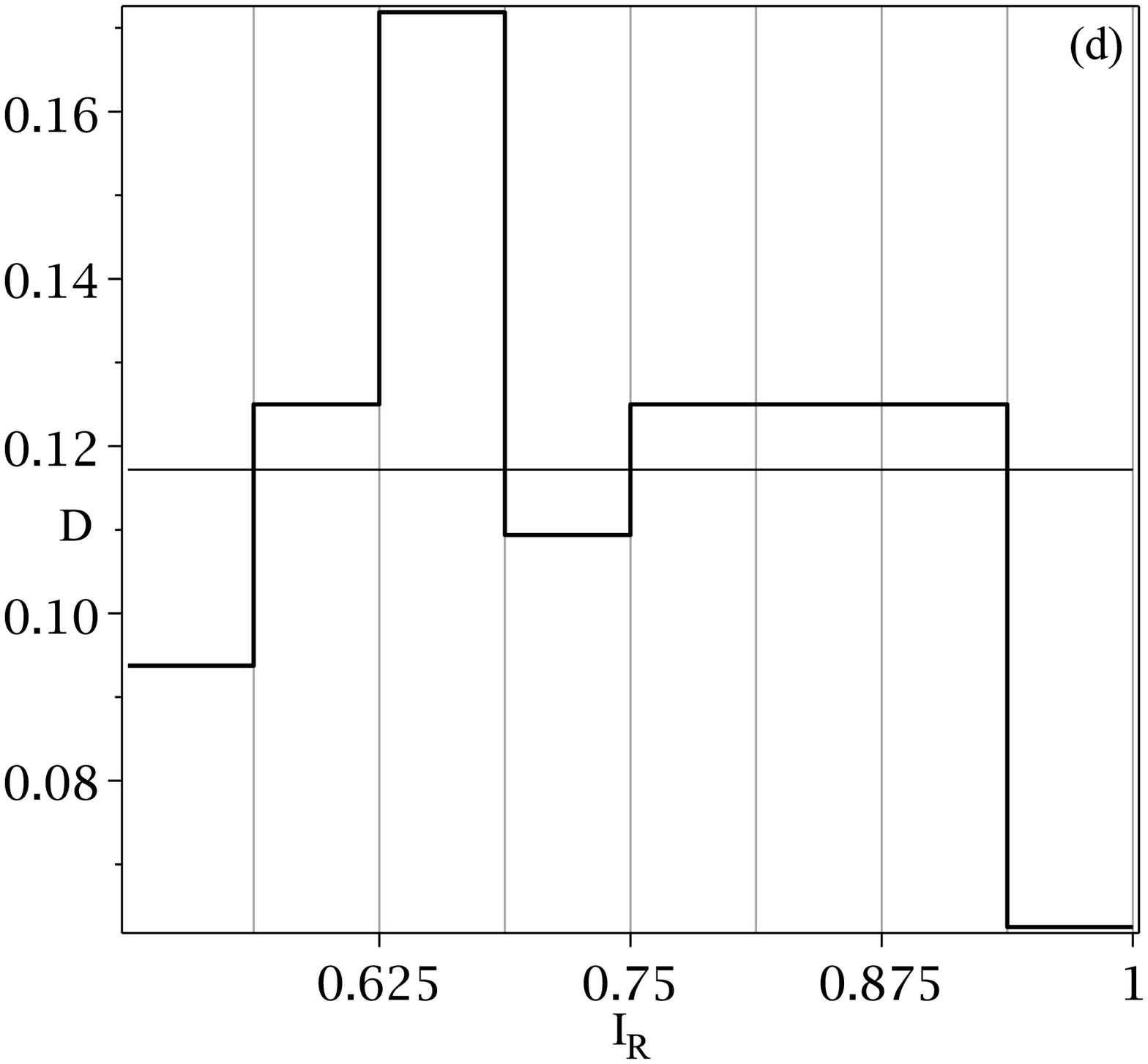} \includegraphics[width=5.9cm]{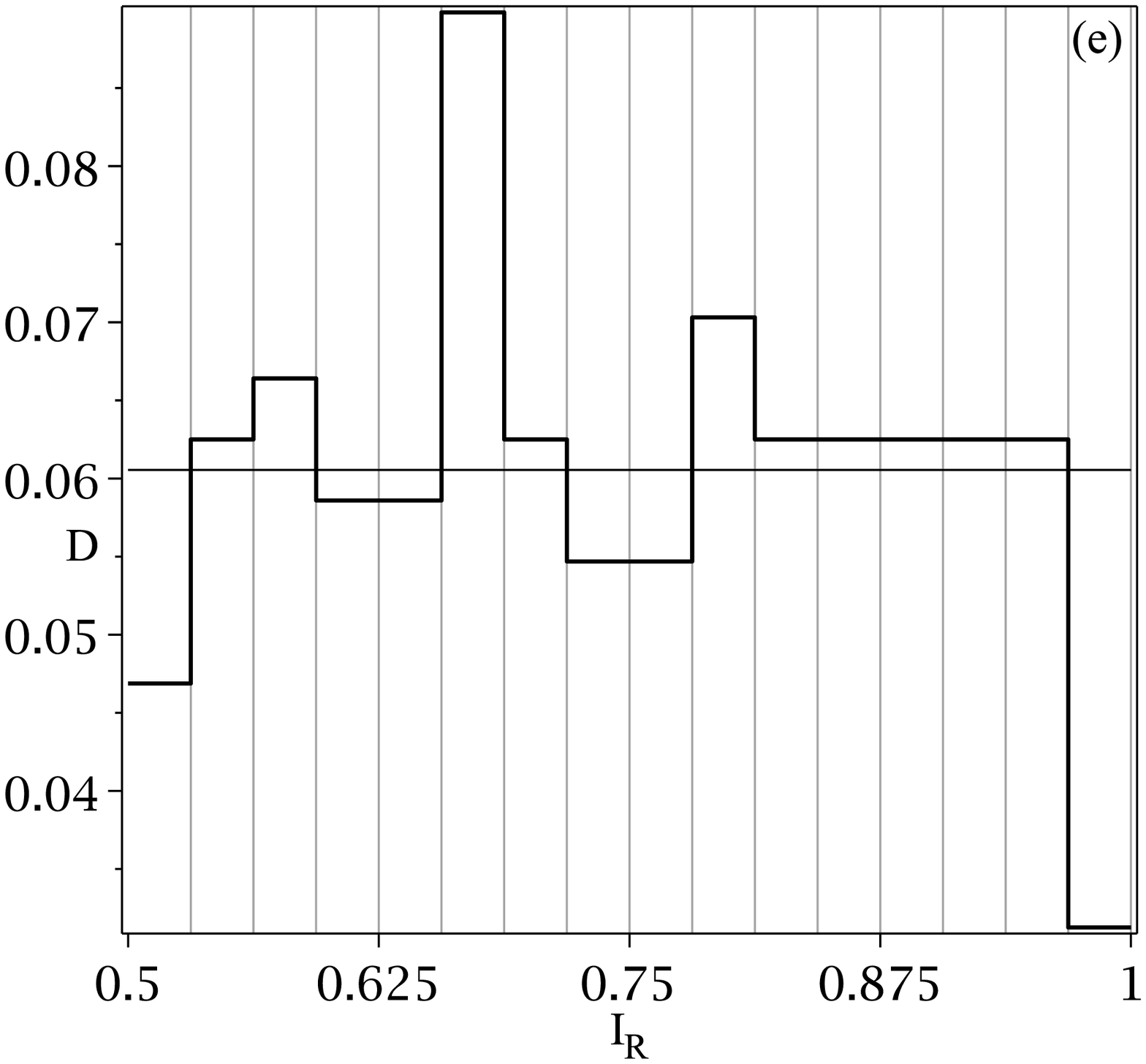} \includegraphics[width=5.9cm]{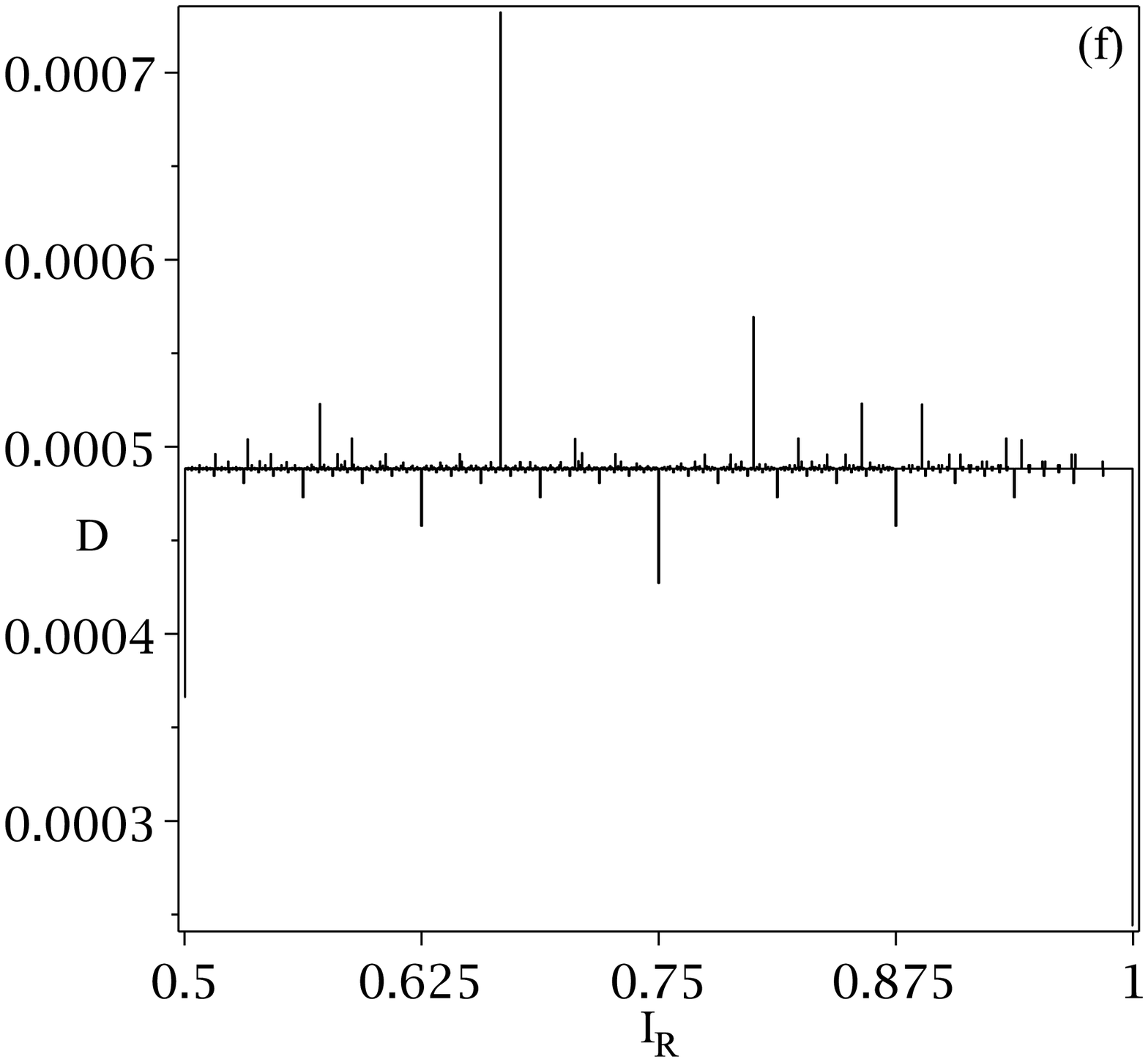}\\
\end{center}
\caption{Non-symmetric holes for a symmetric map and a non-symmetric map: In this figure the
analytically derived diffusion coefficient $D$ is illustrated for
two different dynamical systems as a function of the position of the
hole $I_R$, of size $1/2^{s}$ (cf. Fig.~\ref{fig:6D}) with $I_L$ fixed at $0$. We display the diffusion coefficient
Eq.(\ref{Eq:Diff_ns}) of the doubling map Eq.(\ref{Eq:Bern_lift}) with $s=4,5$ and $12$ in (a), (b) and (c) respectively and the diffusion coefficient
Eq.(\ref{Eq:Diff_tent}) of the tent map Eq.(\ref{Eq:Tent}) with $s=4,5$ and $12$ in (d), (e) and (f) respectively. The horizontal lines give the average value $\left\langle D_s\right\rangle=2(h-h^2)$. }
\label{fig:tentmap}
\end{figure*}
One can also consider the case where the holes are not placed
symmetrically. One way of doing this is by setting $a_1=0$ in
Eq.(\ref{Eq:Bern_lift}) and letting $I_R$ vary independently such that
$0=a_1\leq a_2 \leq 1/2 \leq a_3 \leq a_4 \leq 1$ and
$a_4-a_3=a_2=h=1/2^s$. As before we assume Markov holes, i.e. $a_{3}=1/2 + (i-1) h$ with $i=1,\ldots (2h)^{-1}$. The cumulative function in this setting is
given by
\begin{equation}
T^{ns}(x)=
\left\{
\begin{array}{lc}
\frac{1}{2}T^{ns} (2x )   +x            & \ \ 0\leq x <a_2\\
\frac{1}{2}T^{ns} (2x )   +a_2          & \ \ a_2\leq x <\frac{1}{2}\\
\frac{1}{2}T^{ns} (2x-1 ) +a_2          & \ \ \frac{1}{2}\leq x <a_3\\
\frac{1}{2}T^{ns} (2x-1 ) -x+a_4        & \ \ a_3\leq x <a_4\\
\frac{1}{2}T^{ns} (2x-1 )               & \ \ a_4\leq x \leq 1  \end{array}\right. \:,
\label{Eq:Tfull_nonsymm}
\end{equation}
where the superscript $ns$ refers to the non-symmetrically placed holes.
The diffusion coefficient is then given by
\begin{equation}
               D=T^{ns}(a_2)-T^{ns}(a_4) + T^{ns}(a_3)-h\,,
\label{Eq:Diff_ns}
\end{equation}
cf.\ Eqs.(\ref{Eq:D},\ref{Eq:D2}). The consequence of using
non-symmetrically placed holes can be seen in
Figs.~\ref{fig:tentmap}.(a), (b) and (c), in which the analytical
Eq.(\ref{Eq:Diff_ns}) is evaluated for $s=4,5$ and $12$ respectively
and illustrated as a function of the position of $I_R$. We see that
the step function structure is changed relative to Fig.~\ref{fig:6D}.(c),
(d) and (f). In particular, in Fig.~\ref{fig:tentmap}.(c) we observe a very evenly
distributed set of minima, these being the dyadic rationals which are
in this case all the preimages of $x=0$ in $I_L$ and hence indicative of
increased backscattering. In addition, the average value for the diffusion
coefficient is $\langle D_{s}\rangle= 2(h-h^{2})$ rather than $h$ due to
the constant presence of the running orbit at $x=0$. This follows from
noticing that $T_{i}(a_4)-T_{i+1}(a_3)=-h$, where $T_{i}$ corresponds
to a cumulative function defined by $a_{3}=1/2 + (i-1) h$ with
$i=1,\ldots (2h)^{-1}$. Here, the $a_4$ for $i$ is the same as $a_3$
for $i+1$ and so their itinerary is the same except for the very first
point where $t_{i}(a_4)-t_{i+1}(a_3)=-h$. So we have that the average
over all $i$'s is also equal to $-h$ leaving $\langle D_{s}\rangle=
T(a_2)$ which can be calculated directly to be $T(a_{2})=2(h-h^{2})$.

Accordingly, one can now consider non-symmetric microscopic dynamics. One
possibility for this is to replace the doubling map modulo one in
Eq.(\ref{Eq:Bern}) by the tent map
\begin{equation}
\tilde{\Lambda}(x):[0,1]\rightarrow[0,1]\,,\,
\tilde{\Lambda} (x)=
\left\{
\begin{array}{rl}
2x & 0\leq x <\frac{1}{2}\\
2-2x & \frac{1}{2}\leq x \leq 1\end{array}\right. .
\label{Eq:Tent}
\end{equation}
Now the dynamics no longer commutes with the symmetry $x\to 1-x$.  We again dig two holes with $0\leq a_1 < a_2 \leq 1/2 \leq a_3 < a_4 \leq 1$ into Eq.(\ref{Eq:Tent}) as before and periodically
copy the resulting map over the real line with a lift of degree
one. The cumulative function is given by
\begin{equation}
T_\Lambda (x)=
\left\{
\begin{array}{lc}
\frac{1}{2}T_\Lambda (2x )                 & \ \ 0\leq x <a_1\\
\frac{1}{2}T _\Lambda(2x )   +x-a_1        & \ \ a_1\leq x <a_2\\
\frac{1}{2}T_\Lambda (2x )   +a_2-a_1      & \ \ a_2\leq x <\frac{1}{2}\\
-\frac{1}{2}T_\Lambda (2-2x ) +a_2-a_1     & \ \ \frac{1}{2}\leq x <a_3\\
-\frac{1}{2}T_\Lambda (2-2x ) -x+a_4      & \ \ a_3\leq x <a_4\\
-\frac{1}{2}T_\Lambda (2-2x )              & \ \ a_4\leq x \leq 1  \end{array}\right. \:,
\label{Eq:Tfull_tent}
\end{equation}
where the subscript $\Lambda$ indicates we are considering the
cumulative function for the tent map. The diffusion coefficient is similarly given by
\begin{equation}
               D=T_\Lambda(a_2)-T_\Lambda(a_1)-T_\Lambda(a_4) + T_\Lambda(a_3)-h.
\label{Eq:Diff_tent}
\end{equation}
cf.\ Eq.(\ref{Eq:D}). We can not use the symmetry condition to simplify Eq.(\ref{Eq:Diff_tent})
further as the cumulative function Eq.(\ref{Eq:Tfull_tent}) is not a symmetric function.

Interestingly, if the holes are placed symmetrically such that $a_{4}=1-a_{1}$ and $a_{3} = 1-a_{2}$, then $D=h$ and is independent of the hole position. This follows by noticing that $T_{\Lambda}(a_3)= -1/2 T_{\Lambda}(2a_2)+h$ and $T_{\Lambda}(a_4)= -1/2 T_{\Lambda}(2a_1)$.

If the holes are placed in a non-symmetric way such that
$0=a_1< a_2 \leq 1/2 \leq a_3 < a_4 \leq 1$ and
$a_4-a_3=a_2=h=1/2^s$, then the periodic orbit dependent structure of the diffusion coefficient is re-established. This is illustrated as a function of $I_R$ in Figs.~\ref{fig:tentmap}.(d), (e) and (f) for $s=4,5$ and $12$ respectively. As before, $\langle D_{s}\rangle= T_{\Lambda}(a_2)=2(h-h^{2})$.

A recursive relation of the type of Eq.(\ref{Eq:parent}) can also be established for both doubling and tent maps with holes placed non-symmetrically, as described above. We find that
\begin{eqnarray}
                D_{s} &=& 2 D_{s+1}^{0} + 2 D_{s+1}^{1} - 2^{-s+1},
\label{Eq:parent22}
\end{eqnarray}
where the difference in the last term's exponent is indicative of the
constant running orbit at $x=0$ so that the average rescaled
fluctuations of the child diffusion coefficients are intensified by a
factor of two.

\subsection{Asymptotic behavior}
\label{subsec:asymptotic}

In this subsection we focus our discussion on the first case of symmetric holes in the symmetric map $M(x)$ and comment on its generalizations to the non-symmetric cases towards the end.
We will analyze the behavior of the diffusion
coefficient for the Bernoulli shift as the hole size $h=a_2-a_1$ goes
to zero. By doing this, the hole will converge to a point which could
be a running orbit, a standing orbit, or a non-periodic orbit.  We
derive equations which give the asymptotic behavior in all three cases
and use them to obtain the diffusion coefficient in terms of all
periodic orbits in a hole of small but finite size. In order to do
this, we first rewrite the $(T(a_2)-T(a_1))$ term from
Eq.(\ref{Eq:D2}) with Eq.(\ref{Eq:Tsum}) to
\begin{eqnarray}\nonumber
               T(a_2)-T(a_1) =\ \ \ \ \ \ \ \ \ \ \ \ \ \ \ \ \ \ \ \ \\
               \lim_{n \rightarrow \infty} \sum_{k=0}^n\frac{1}{2^k}\left(t\left(\tilde{M}^k\left(a_2\right)\right)-t\left(\tilde{M}^k\left(a_1\right)\right)\right).
\label{Eq:Ta2-Ta1}
\end{eqnarray}
First consider the case that $I_L$ converges to a running orbit, that
is, a periodic point $x_p$ of period $p$, which does not enter $I_R$ under
forward iteration. In this case, from Eq.(\ref{Eq:tfull}) we see that the
only contributions to Eq.(\ref{Eq:Ta2-Ta1}) come when $k=lp, l \in \nz$
\begin{eqnarray}\nonumber
                 T(a_2)-T(a_1)&\sim& \lim_{n \rightarrow \infty} h\left( \sum_{l=0}^n \frac{1}{2^{lp}}\right) \ \ \left(h\rightarrow 0\right) \\
                              &=&  h\left(\frac{1}{1-2^{-p}}\right) \ \ \left(h\rightarrow 0\right).
\label{Eq:Ta2-Ta1nbs}
\end{eqnarray}
Evaluating Eq.(\ref{Eq:D2}) with Eq.(\ref{Eq:Ta2-Ta1nbs}) we get
\begin{equation}
               D(x_p)\sim h J_{p}^{r}=h\left( \frac{1+2^{-p}}{1-2^{-p}}\right) \ \ \left(h\rightarrow 0\right),
\label{Eq:DLimitnbs}
\end{equation}
where the superscript $r$ denotes a running orbit.  Now consider the
case where $I_L$ converges to a standing orbit, that is, a periodic
point $x_p$ of period $p$, which enters $I_R$ under forward
iteration. Note that due to the symmetry of the holes this will always
occur at time $p/2$ and hence standing orbits always have even
periods. In this case we get a positive contribution to
Eq.(\ref{Eq:Ta2-Ta1}) when $k=lp$, and a negative contribution when
$k=lp/2$,
\begin{eqnarray}\nonumber
                 T(a_2)-T(a_1)&=& \lim_{n \rightarrow \infty} h\left( \sum_{l=0}^n \frac{(-1)^l}{2^{\frac{lp}{2}}}\right) \ \ \left(h\rightarrow 0\right) \\
                              &\sim&  h\left( \frac{1}{1+2^{-\frac{p}{2}}}\right) \ \ \left(h\rightarrow 0\right).
\label{Eq:Ta2-Ta1bs}
\end{eqnarray}
In this case Eq.(\ref{Eq:D2}) evaluates as
\begin{equation}
               D(x_p)\sim h J_{p}^{s}=h \left(\frac{1-2^{-\frac{p}{2}}}{1+2^{-\frac{p}{2}}}\right) \ \ \left(h\rightarrow 0\right).
\label{Eq:DLimitbs}
\end{equation}
The final case to consider is where $I_L$ converges to a point which
is non-periodic. In this setting the only contribution to Eq.(\ref{Eq:Ta2-Ta1})
 comes from the $k=0$ term and therefore
\begin{equation}
               D\sim J_{p}^nh=h   \ \ \left(h\rightarrow 0\right),
\label{Eq:Dnonp}
\end{equation}
 which reproduces the simple random walk result. In summary, we have
\begin{equation}\label{Eq:Dcomb}
 D\sim J_p^\wp h=\left\{\begin{array}{cc}
 h\frac{1+2^{-p}}{1-2^{-p}}&\wp=r\\
 h\frac{1-2^{-p/2}}{1+2^{-p/2}}&\wp=s\\
 h&\wp=n
\end{array}\right. .
\end{equation}
Eq.(\ref{Eq:Dcomb}) gives us a good explanation for the structure that
we see in Fig.~\ref{fig:6D} with improved agreement for small holes
(large $s$). As $s$ is increased, the different asymptotic regimes can
be seen in the step function. For example, when $I_L$ is placed on
a running orbit such as $x=0$ where $p=1$, Eq.(\ref{Eq:DLimitnbs}) tells
us that $D=3h$ for small $h$. When $I_L$ is placed on a standing orbit
like $x=1/3$ with $p=2$, Eq.(\ref{Eq:DLimitbs}) tells us that $D=h/3$
for small $h$. These deviations from the average value of $h$ are
observed in Fig.~\ref{fig:6D}.

\begin{figure*}[htb]
\includegraphics[width=8.5cm]{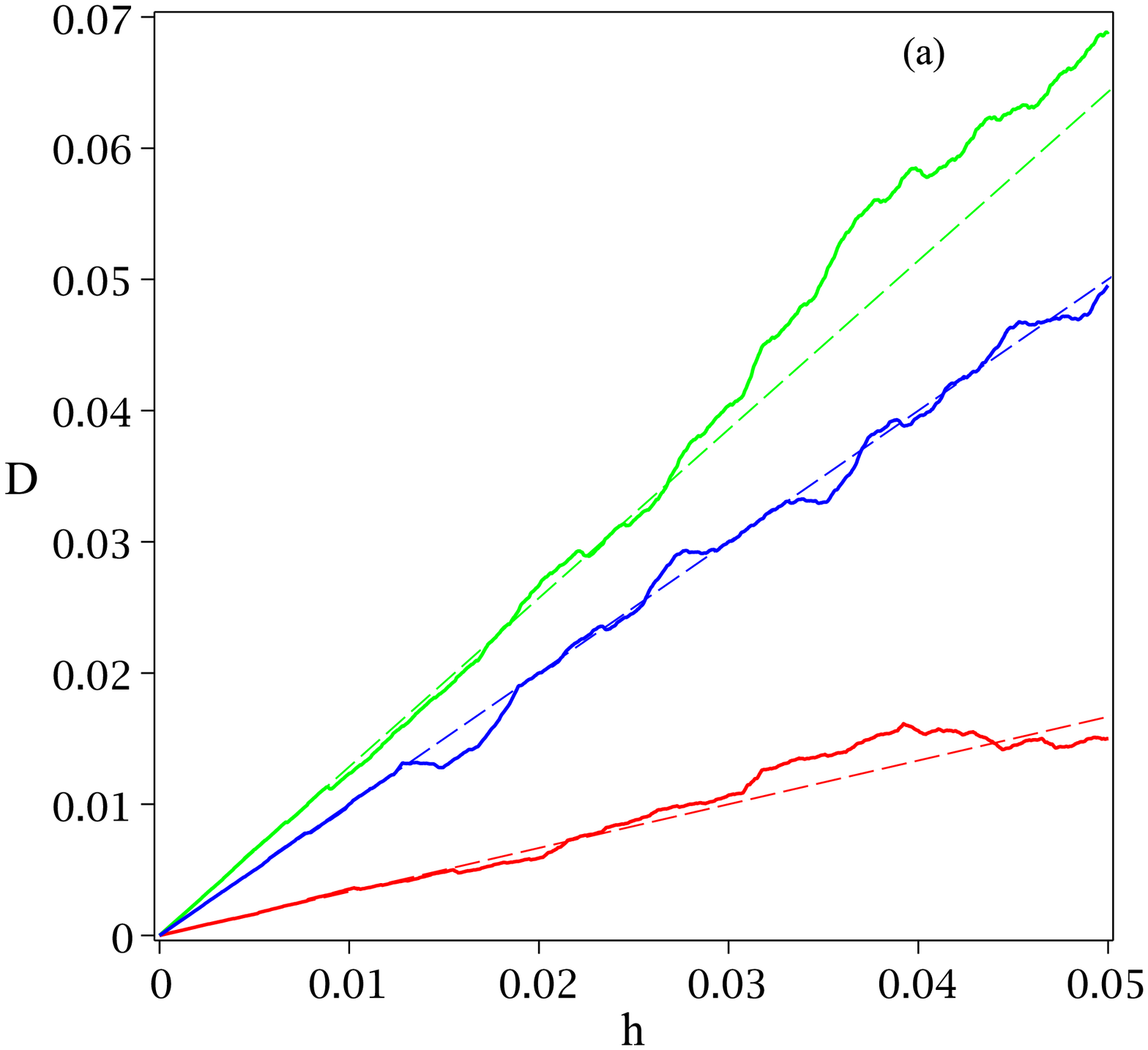} \includegraphics[width=8.5cm]{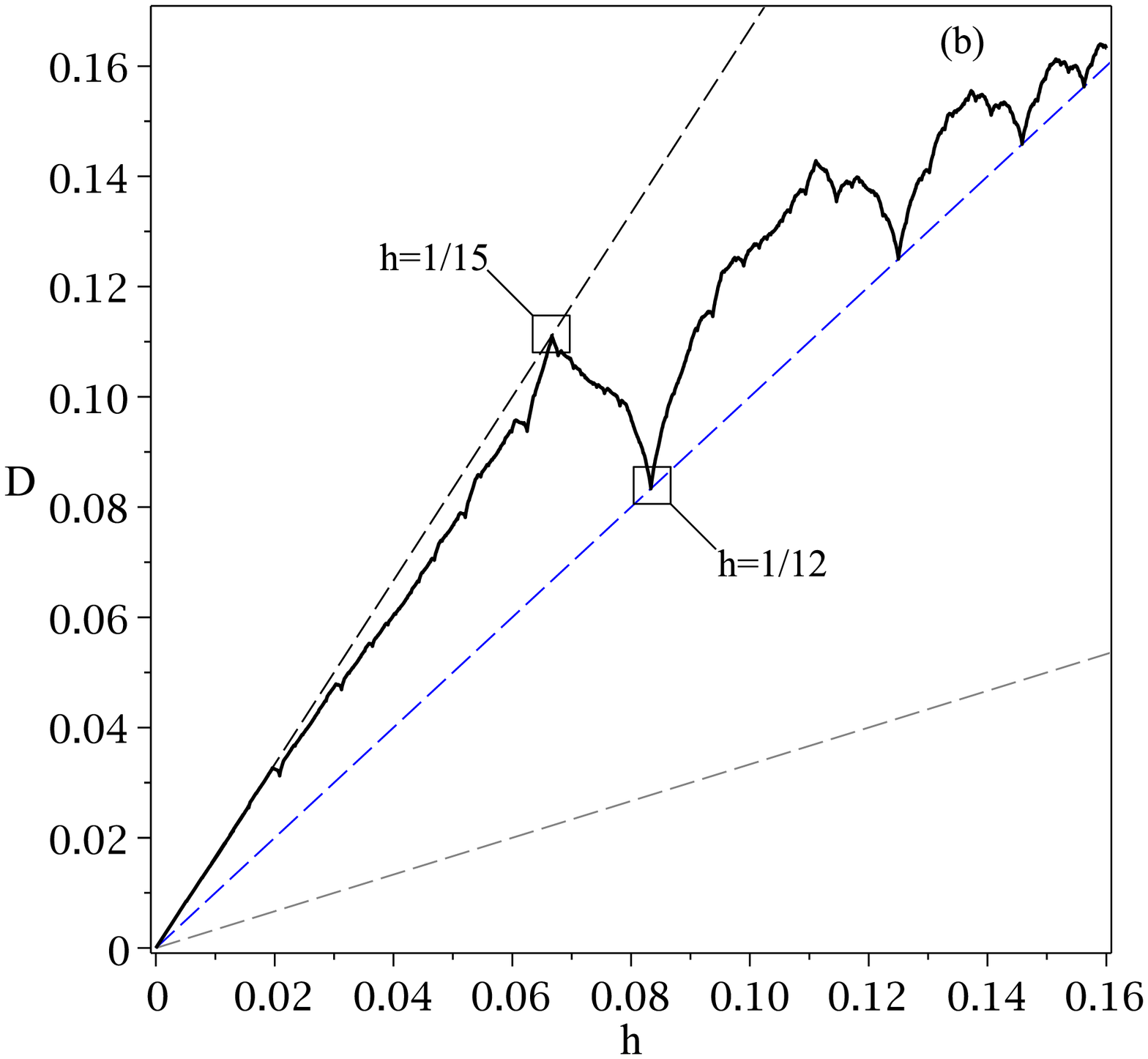}
\caption{(Color online) The asymptotic regimes: In (a) the diffusion
coefficient $D(h)$ for holes centered on three different
classes of points in $M(x)$ is illustrated as a function of
the hole size $h$ (from top to bottom); $x=1/3$ a standing
orbit (red), $x=\sqrt{5}/2-17/25$ a non-periodic orbit (blue) and
$x=1/7$ a running orbit (green) along with the different asymptotic
regimes $h/3$, $h$ and $9h/7$ respectively shown by dashed
lines. These asymptotic regimes correspond to the result of
Eq.(\ref{Eq:Dcomb}). In (b), the position of the left
boundary of the hole $a_1=1/3$ is fixed and $h$ is again decreased
continuously. We observe that $x=1/3$ becomes a running orbit when a
critical point, and the asymptotic regime of $5h/3$ (black dashed,
top) illustrates this. The line $h/3$ (grey dashed, bottom) is what
one would expect if $x=1/3$ was contained in $I_L$ yielding a
standing orbit, and the random walk solution is given by the blue
dashed line (middle). The two symbols (squares) identify parameter
values where the right boundary point $a_2$ of the hole generates a standing
orbit, respectively a running orbit.}
\label{fig:asymp}
\end{figure*}

A further consequence of Eq.(\ref{Eq:Dcomb}) is the intriguing result
that for small hole size, one can not rely on the simple random walk
approximation for an accurate description of the diffusion coefficient
\cite{SFK,GF2,RKdiss,dcrc}. Rather, one must go beyond this theory and
take into account the periodic orbit structure of the system, and in
particular, the periodic orbits contained in the holes. The
asymptotic regime that one obtains for small $h$ will be dependent
upon the type of point that the holes converge to. The authors
are aware of only one other published result on a one-dimensional
system in which the random walk approximation theory is violated
\cite{Kni11}. In this case the phenomenon was explained in terms of
ergodicity breaking, which is not the case here.

We can now go beyond the small hole limit by combining the above
results with the parent-child hole relation of
Eq.(\ref{Eq:parent}). For large $n$ we have that
\begin{equation}
D_{s+n}= 2^{-s-n} J_{p}^{\wp} \ \ \left(n\rightarrow \infty\right),
\label{Eq:nlim}
\end{equation}
with $\wp\in\{r,s,n\}$ depending on the limiting point of the hole as
in Eq.(\ref{Eq:Dcomb}). Hence, we may now express $D_{s}$ in terms of
all periodic orbits of period $p$ which intersect the holes,
\begin{equation}
D_{s}= 2^{-s}\left( 1 + \sum_{p} (J_{p}^{\wp}-1) \right) \ \ \left(n\rightarrow \infty\right).
\label{Eq:poexp}
\end{equation}
Note that a periodic orbit that intersects the parent hole more than
once just gets added each time. Also, as discussed below, all periodic
orbits are counted as running if they occur at the end of the interval.
Eq.(\ref{Eq:poexp}) suggests that the observed fluctuations of $D_{s}$
from its average $\langle D_{s} \rangle=2^{-s}$ are due to the individual
fluctuations of the infinitely many periodic orbits which intersect the holes.

As expected from periodic orbit theory \cite{CAMTV01}, a very large
number of periodic orbits is needed to trace the hole
accurately. However, if the periodic orbits are ordered appropriately
the sum may be truncated to produce good approximations to $D_{s}$
\cite{DettMor97}. Note that the optimal ordering (for fast
convergence) of periodic orbits is by a modified version of the length
of the orbit; $J_{p_{1}}^{r}\approx J_{p_{2}}^{s}$ for $2p_{1}\approx
p_{2}$. In other words backscattering is much more dominant for orbits
of equal period (see Fig.~\ref{fig:asymp}).

We can further study the asymptotic behavior for the three different
cases derived above and the finite hole size result by
reducing $I_L$ continuously. Fig.~\ref{fig:asymp}(a) illustrates these
different regimes.

When using Eq.(\ref{Eq:Dcomb}), care needs to be taken when $I_L$
converges to a point from the left or the right, i.e., it is not
centered on a point and reduced in size. In this case a boundary point
of the hole, $a_1$ or $a_2$, is kept fixed. If the boundary
point is periodic, points near it in the interior of $I_L$ miss $I_R$,
and so it is always a running orbit. For example, as illustrated in
Fig.~\ref{fig:6D} and Fig.~\ref{fig:asymp}.(a), $x=1/3$ is a period two
standing orbit when $1/3$ is in the interior of $I_L$ and the
asymptotic regime for small $h$ when $I_L$ converges to $1/3$ is given
by Eq.(\ref{Eq:DLimitbs}) as $h/3$. However, if $a_1=1/3$ is fixed and
$h$ goes to zero, we must use Eq.(\ref{Eq:DLimitnbs}) to evaluate the
asymptotic regime as in this case $1/3$ is a running
orbit. Eq.(\ref{Eq:DLimitnbs}) tells us that the asymptotic regime is
in fact $5h/3$. This additional topological subtlety that must be
considered is illustrated in Fig.~\ref{fig:asymp}.(b).

We close this subsection by considering small hole approximations to the non-symmetric models considered in Sec.\ref{Subsec:other}. In analogy to Eq.(\ref{Eq:Dcomb}) for the doubling map with non-symmetric holes as in Eq.(\ref{Eq:Tfull_nonsymm}) and (\ref{Eq:Diff_ns}), we find through similar arguments that
\begin{equation}
 J_p^r=  1+\frac{1}{1-2^{-p}} ,
\label{Eq:j1}
\end{equation}
if $I_{R}$ is converging onto a running periodic orbit of period $p$ (since $T(a_{2})= 2 h$ as $a_{2}\rightarrow 0$). In contrast with the symmetric holes, here all backscattering is due to dyadic rationals. This is evident in Fig.~\ref{fig:tentmap}. Hence, we find
\begin{equation}
 J_p^s= 2-2^{1-n},
\label{Eq:j2}
\end{equation}
if $I_{R}$ is converging onto a dyadic rational of the form $i/2^{n}$ where $i$ is some positive integer. A non-periodic point gives
\begin{equation}
 J_p^n= 2.
\label{Eq:j3}
\end{equation}

For the tent map with non-symmetric holes as considered in Sec.\ref{Subsec:other}, Eqs.(\ref{Eq:j1}) and (\ref{Eq:j3})
carry over, while
\begin{equation}
 J_p^s= 2-2^{-n},
\label{jj2}\end{equation}
since dyadic rationals need one more forward iteration of the tent map to enter $I_L$. Finally, the periodic orbit sum of Eq.(\ref{Eq:poexp}) modified by a factor of $2$ also holds for the non-symmetric models considered here.

\section{The escape rate}
\label{sec:escrate}

It is interesting to note that an analytical relationship between the escape rate of a
spatially extended diffusive dynamical system with absorbing
regions and its diffusion coefficient has been established
by the escape rate theory of diffusion \cite{GN,Gasp,RKdiss,Do99,Kla06}. Motivated by
Refs.\onlinecite{Bun11} where the complicated dependence of the
escape rate on position and size of a hole has been studied, here we
focus on the relationship between the open map $\tilde{M}(x)$ on the
unit interval with the symmetric holes $I_L$ and $I_R$ serving as absorbing
regions and the diffusion coefficient of the corresponding coupled,
spatially extended system.  That is, for calculating the escape rate
any orbit that enters either of these intervals is removed from the
system, and in this way points from an initial density escape, while
for calculating the diffusion coefficient all points remain within the
system by performing `jumps' when hitting these intervals, as defined
by the lift Eq.(\ref{Eq:lift}). An interesting question is to which
extent the coupled diffusive `jump dynamics' of the spatially extended
system is already captured by the escape rate of the interval map that
defines the unit cell of this lattice.

The main result from Ref.\onlinecite{Bun11}
concerning the escape rate is that escape will occur fastest through a
hole whose minimal period is highest, or equivalently, the escape rate
will be slowest through the hole which has the smallest minimal
period. By minimal period we mean the smallest period of all the
periodic points in a hole.

\begin{figure}[htb]
\begin{center}
\includegraphics[width=8.5cm]{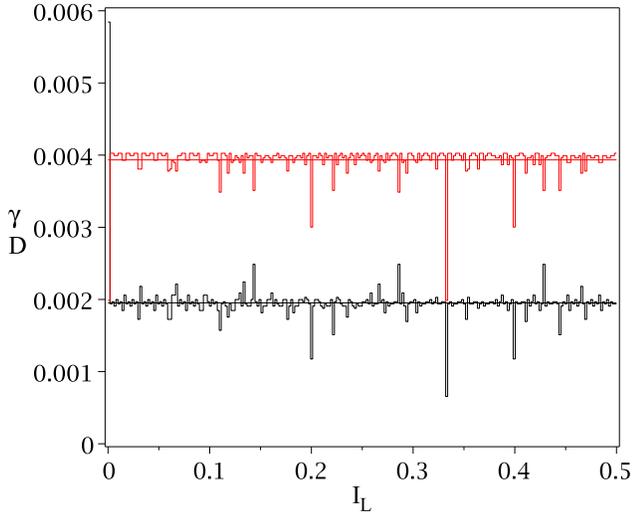}
\end{center}
\caption{(Color online) Comparing the diffusion coefficient with the
escape rate: In this figure the diffusion coefficient for the
doubling map is illustrated in black (bottom) alongside the escape
rate of the corresponding open system in red (top) as a function of
the hole $I_L$. The thin horizontal lines illustrate the
average value to aid visual comparison of the fluctuations:
$\left\langle D\right\rangle=1/2^9$ and $\left\langle \gamma \right\rangle\simeq 0.00393$ (3s.f.) $\simeq 1/2^8$.
There is a clear relationship between the
structure of these functions although intervals which give relatively
high diffusion coefficients will give relatively low escape rates. The
calculation has been performed for intervals $I_L$ of size
$1/2^9$.}
\label{fig:D_E_comp}
\end{figure}

In order to calculate the escape rate of our system we look at the
transition matrix induced by the dynamics. The escape rate $\gamma$
can be evaluated via the largest eigenvalue $\nu$ of this transfer
matrix \cite{RKD,RKdiss,GaKl,Gasp,Kla06}
\begin{equation}
                \gamma= -\ln \nu.
\label{Eq:esc_rate}
\end{equation}
In Fig.~\ref{fig:D_E_comp} solutions to Eq.(\ref{Eq:esc_rate}) are
illustrated for $s=9$ and compared with the diffusion coefficient in
the corresponding extended system. Here we see that similar structures
arise with deviations from the average occurring for both phenomena on
the same intervals. In order to quantify these deviations we can
compare Eq.(\ref{Eq:Dcomb}) with Theorem $4.6.1$ from
Ref.\onlinecite{Bun11}, which generalized to Theorem
$2.1$ from Ref.\onlinecite{KeLi09}. It gives the escape rate for
small hole size in the doubling map with one hole and can
easily be generalized to escape through two holes as is the case
here. This theorem states that the escape rate for small $h$ with a
running orbit (no iterate of the orbit reaches the second hole) is
given by
\begin{equation}
                \frac{\gamma(x_p)}{h}\to 2\left(1-\frac{1}{2^p}\right) \ \ \left(h\to 0\right),
\label{Eq:escp}
\end{equation}
where $x_p$ is the lowest period point in the hole with period $p$.
For a standing
orbit (the periodic orbit is in both holes) the period is effectively
halved and we get
\begin{equation}
                \frac{\gamma(x_p)}{h}\to2\left(1-\frac{1}{2^{p/2}}\right) \ \ \left(h\to 0\right).
\end{equation}
When the hole converges to a non-periodic point, the theorem
 states that the escape rate is given by
\begin{equation}
                \frac{\gamma(x)}{h}\to2 \ \ \left(h\to 0\right).
\label{Eq:escnonp}
\end{equation}
From Eq.(\ref{Eq:escp}) the relative deviation from the average escape
rate $\langle \gamma \rangle =2h$ is given by
\begin{equation}
               \gamma(x_p)-\langle \gamma \rangle =-\frac{2h}{2^p}.
\label{Eq:dev_esc_run}
\end{equation}
for a running orbit and
\begin{equation}
               \gamma(x_p)-\langle \gamma \rangle =-\frac{2h}{2^{p/2}}.
\label{Eq:dev_esc_sta}
\end{equation}
for a standing orbit.
Similarly, the relative deviation from the average diffusion coefficient
 $\langle D \rangle =h$, for a running orbit, can be obtained from Eq.(\ref{Eq:DLimitnbs}) as
\begin{equation}
               D(x_p)-\langle D \rangle =\frac{2h}{2^p-1},
\label{Eq:dev_Dnbs}
\end{equation}
whilst for standing orbits, via Eq.(\ref{Eq:DLimitbs}), the relative deviation is given by
\begin{equation}
               D(x_p)-\langle D \rangle =-\frac{2h}{2^{p/2}+1}.
\label{Eq:dev_Dbs}
\end{equation}
Eqs.(\ref{Eq:dev_esc_run}), (\ref{Eq:dev_esc_sta}), (\ref{Eq:dev_Dnbs}) and (\ref{Eq:dev_Dbs})
help us explore the relationship between the diffusion coefficient of
the extended system with the escape rate of the open system. An
obvious difference is the absence of backscattering in the escape
rate. However a more striking one is that the average escape rate does
not equal the algebraic mean of all escape rates as for diffusion
coefficients. That is $\langle \gamma \rangle \neq
\frac{1}{2^{s-1}}\sum_{j=1}^{2^{s-1}} \gamma_{s}^{j}$ which is obvious
from Fig.~\ref{fig:D_E_comp} but is also suggested by
Eq.(\ref{Eq:escp}). The two symmetric holes are coupled differently
for the escape problem and in a much more complicated way than as in
Eq.(\ref{Eq:parent}) by involving the eigenvalues of $2^{s-1}\times
2^{s-1}$ transfer matrices. However, for small holes this coupling
decays rapidly revealing the similarities which are seen in
Fig.~\ref{fig:D_E_comp}.
We remark that while for non-symmetric holes as the ones considered in Sec.\ref{Subsec:other} the diffusion coefficients may display qualitative differences (e.g. no position dependence), the corresponding escape rates differ only quantitatively.

\section{Conclusion}
\label{sec:conclusion}

The aim of this paper was to study the `dependence of chaotic
diffusion on the size and position of holes'. The answer was
provided by analytically deriving the diffusion coefficient for
both a symmetric and a non-symmetric one-dimensional piecewise linear
map as a function of the size and position of a hole. We showed that
for both maps the diffusion coefficient is a complicated function
of the position and a non-monotonic function of the size of the holes,
despite the fact that the underlying reduced dynamics is not changed,
as is the case in previously studied models \cite{RKD,GaKl,Kla06,Kni11}.
This finding implies that, surprisingly, making a hole smaller can increase the diffusion coefficient. These results we explained
via the periodic orbit structure of the map and the ideas of running
and standing periodic orbits.

We furthermore found that the asymptotic regime that one obtains for
small hole size is a function of the type of periodic orbit that the
holes converge to. This is another important result, since it
generalizes the standard uncorrelated random walk approximation of
simple stochastic processes. It implies that this random walk
approximation may not always give accurate estimates for the diffusion
coefficient of a chaotic dynamical system. We have also obtained a
new expansion for the diffusion coefficient of finite size holes in
terms of periodic orbits and discussed their relative importance for
the dynamics. In our setting, a periodic orbit can either be a running
or a standing orbit. The presence of a standing orbit has the effect
of reducing the diffusion coefficient relative to the average value
whilst the presence of a running orbit has the effect of increasing it
relative to the average value.

We finally numerically calculated the escape rate of the corresponding
open system and compared it with the diffusion coefficient
thus relating diffusion and escape in a new manner. We found
that the diffusion coefficient and escape rate are both dependent upon
the underlying periodic orbit structure of the map, although
differences arise which we explain as a difference in
the coupling between holes.

An interesting open question is whether there exists a parent-child scaling relation
for the escape rate that is similar to the scaling relation for the
diffusion coefficients Eqs.(\ref{Eq:parent},\ref{Eq:parent22}). Another interesting open question is to which extent the above effects can
be observed in computer simulations of diffusion in higher
dimensional, more physically realistic systems such as suitably
adapted periodic Lorentz gases \cite{Kla06,Gasp} and related particle
billiards \cite{HaGa01,HaKlGa02}. This should pave the way to
design experiments where these effects might be observable, such as
modified cold atom experiments on atom-optics billiards
\cite{Rai01,FKCD01}.


\end{document}